\documentclass[12pt]{iopart}
\usepackage{amssymb}
\usepackage{siunitx}
\usepackage{graphicx}
\usepackage{booktabs}
\usepackage{subcaption}
\usepackage{color}

\usepackage[pdftex]{hyperref} 
\graphicspath{{./figures/}}
\newcommand{\myEmph}[1]{#1}
\newcommand{\myEmphTwo}[1]{#1}
\newcommand{\myEmphThree}[1]{#1}

\begin{document}
\title[Multiphysics simulation of aspherical piezo-glass membrane lenses]{\myEmph{Multiphysics simulation of the aspherical deformation of piezo-glass membrane lenses including hysteresis, fabrication and non-linear effects}}
\author{Florian Lemke$^1$, Yasmina Frey$^1$, Binal P Bruno$^1$, Katrin Philipp$^2$, Nektarios Koukourakis$^2$, J\"urgen Czarske$^2$, Ulrike Wallrabe$^1$ and Matthias C Wapler$^1$\\
}
\address{$^1$Laboratory for Microactuators, Department of Microsystems Engineering (IMTEK),
University of Freiburg, Georges-K\"ohler-Allee 102, 79110 Freiburg, Germany\\
$^2$Chair of Measurement and Sensor System, Institute of Principles of Electrical and Electronic Engineering, Technical University of Dresden, Helmholtzstra{\ss}e 18, 01069 Dresden, Germany}
\ead{wallrabe@imtek.uni-freiburg.de}
\vspace{10pt}
\begin{indented}
\item[]March 2019
\end{indented}
\begin{abstract}
In this paper we present and verify the non-linear simulation of an aspherical adaptive lens based on a piezo-glass sandwich membrane with combined bending and buckling actuation. To predict the full non-linear piezoelectric behavior, we measured the non-linear charge coefficient, hysteresis and creep effects of the piezo material and inserted them into the FEM model using a virtual electric field. We further included and discussed the fabrication parameters -- glue layers and thermal stress -- and their variations. To verify our simulations, we fabricated and measured a set of lenses with different geometries, where we found good agreement and show that their qualitative behavior is also well described by a simple analytical model. We finally discuss the effects of the geometry on the electric response and find, e.g., an increased focal power range from $\pm4.5$ to $\pm\SI{9}{\per\meter}$ when changing the aperture from \SI{14}{} to \SI{10}{\milli\meter}.
\end{abstract}
\vspace{2pc}
%
\submitto{\SMS}
%
%
%
\section{Introduction}
Fluidic varifocal lenses provide a fast and efficient method to change the focal power in optical beam paths without mechanical movement, e.g.~as used in novel adaptive scanning microscopes~\cite{Koukourakis:14,HiLo} or flow velocimetry~\cite{Velocity}. 
There are two main types of adaptive lenses, that use different physical principles. The first class of lenses uses a controlled change of the refractive index, such as lenses based on liquid crystals~\cite{LC1,LC2} or acoustic pressure gradients~\cite{Mermillod-Blondin:08}. The second type of lenses uses a change of the curvature between two media with different refractive index. Examples for this kind are glass- and polymer-membrane fluid lenses~\cite{ESmembrane,Schneider:09,PoLight,Bonora:15}, lenses based on dielectric electroactive polymers ~\cite{doi:10.1002/adfm.201403942} and electrowetting lenses~\cite{Berge2000, Kopp}. Each kind of lens type has its own advantages and disadvantages: Electrowetting lenses generally provide a very low wavefront error, but they are relatively slow~\cite{Berge2000}. Lenses with a polymer membrane and integrated actuation can offer a relatively compact setup but are yet not very fast, because of their soft membrane and a strong fluidic damping~\cite{0960-1317-19-9-095013}. Furthermore, correcting not only the defocus, but also spherical aberrations, requires a second device within the optical path/setup for these kind of lenses, e.g. deformable mirrors based on piezoelectric~\cite{Shaw:10} or electrostatic actuation\cite{doi:10.1046/j.1365-2818.2002.01004.x} or liquid crystal spatial light modulators~\cite{doi:10.1002/lpor.200900047}. 

The main advantage of our design~\cite{Matthias14} is that we are able to control both, the defocus and the spherical aberrations simultaneously. \myEmph{In~\cite{Katrin17,MatthiasSpiegel} we demonstrated experimentally that we can independently control the focal power and aspherical coefficient and also showed the achievable aspherical and focal tuning range for one of our lenses~\cite{Katrin17}. We used this ability in~\cite{Katrin18} to correct spherical aberrations in a confocal microscope at different focal depths inside a sample.}
Using only one transmissive device instead of multiple transmissive and reflective devices, we are able to simplify the optical setup and make it more compact. The active element of our lens concept is an active piezo-glass sandwich membrane, where an ultra thin glass membrane is glued in-between two piezo rings, which directly deforms the membrane and hence changes the focal length due to a transparent fluid (or polymer~\cite{TRANSD}) that is added below the membrane. Other piezo actuated glass membrane lenses, e.g. \cite{PoLight}, can also change their focal length, but have only one degree of freedom and hence cannot correct aberrations.

In~\cite{Matthias14} we presented the basic concept of this lens concept with high resonance frequencies ($>$1~kHz) and a large aperture (\SI{12}{\milli\meter} clear aperture vs. \SI{18}{\milli\meter} diameter). It achieved a focal power range of approximately $\pm\SI{4}{\per\meter}$ for a very compact design with the additional ability to tune the spherical behavior. In~\cite{Actuator16,Actuator18} we increased the focal power range for the same geometric dimensions and materials to more than $\pm \SI{6}{\per\meter}$ by a modification of the actuators using an in-plane polarization and an induced pre-stress in the fabrication process. 

The aspherical behavior can be adjusted using two different actuation modes: On the one hand, the "bending mode", with one contracted and one expanded piezo ring, leads to a rather spherical deformation of the glass membrane. On the other hand the so called "buckling mode", where both piezos contract, results in a more \myEmph{hyperbolic} shape. Compared to polymer membrane fluid lenses, the stiff glass membrane of our lens results in a short response time below 0.2 ms~\cite{MatthiasMEMS}.

A similar configuration was later also used in~\cite{Bonora:15} where the authors actuated a glass membrane with segmented piezo actuators to control not only spherical, but also higher order aberrations such as astigmatism and coma. As the main purpose of this lens is the higher order aberration correction, they achieve only a focal power range of approximately $\pm \SI{0.5}{\per\meter}$ and and operating frequencies of \SI{200}{\hertz} \cite{Bonora18}.  Similarly, their setup of using two glass membranes, two sets of actuators and an additional stiffening glass window is more bulky and complex than our configuration. There are also electrowetting lenses that allow correction of higher order aberration, e.g. astigmatism~\cite{Kopp}, but as their contact angle is the only degree of freedom, they are not able to directly control the spherical aberrations independently from the focal power.

To fully understand the actuation principle, design lenses with desired properties or optimize the working range of the focal power and spherical tuning of the present lens, it is essential to predict the expected surface deformation of the actuated glass membrane. Hence, in this paper, we develop a numerical simulation using \textit{COMSOL Multiphysics} to predict the deformation of the lens profile as a function of the applied electric signal. In~\cite{Schneider2008} the authors simulated the focal power and spherical deformation of a an adaptive lens with a passive polymer membrane based on fluidic displacement by piezo actuators. In~\cite{NGUYEN201759} the effects of different piezo electrode configurations on the glass membrane deformation of the adaptive polymer lens of~\cite{PoLight} were simulated. While these papers consider a linear system with a single degree of freedom, our combined bending and buckling actuation is intrinsically a non-linear effect with two degrees of freedom, and we will also take into account the effects of the fabrication and of the non-linear response of the piezo material.

The same actuator with a metal membrane instead of glass membrane was later also used in~\cite{Woias} for a micro pump system. The authors of that paper measured the radial contraction of the piezo rings and from that value predicted the center point deflection of the membrane. In contrast, we measured and modeled the full non-linear piezoelectric response of the material and studied not only the center deflection, but the entire surface deformation of the membrane.
There exist approaches to simulate the non-linear effects of piezo materials, e.g. hysteresis~\cite{COMSOLHYS} based on the  models of~\cite{LANDIS}, using weak form PDEs for implementation in \textit{COMSOL}. In addition to being less trivial to implement, they also require a precise knowledge of different material properties. Our technique, however, provides a fast and simple way to implement the full non-linear behavior for a specific voltage function using only a single material characterization measurement.

In this paper we first describe the operating principle of the lens in detail in section~\ref{sec:op}, where we also show analytic approaches to approximate the voltage-dependent focal power as a function of the geometric parameters. In section~\ref{sec:sim} we explain, how we set up a simulation and address issues such as hysteresis. Then, we describe the fabrication process and the measurements of the lens prototypes in sections~\ref{sec:fab} and~\ref{sec:meas}. In section~\ref{sec:eval}, we compare the results of the simulation to the measurement data and analyze effects of fabrication tolerances. We finally conclude our results in section~\ref{sec:res}.

\section{Operating principle and analytic approximation} \label{sec:op}
\begin{figure*}[bhpt!]
	\centering
	\def\svgwidth{0.8\columnwidth}
	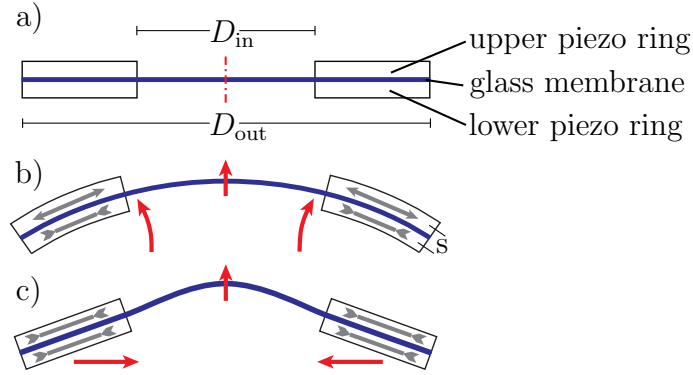
	\caption{Schematic cross-section through the active part of the lens for (a)~non-actuated state, (b)~"bending mode" with the upper piezo ring expanded and the lower piezo ring contracted and (c)~"buckling mode" with both piezos contracted.~\cite{Matthias14}}
	\label{fig:BendingAndBuckling}
\end{figure*}

The active part of the lens consists of two out-of-plane polarized piezo-rings that are bonded to an ultra thin glass membrane as shown in \fref{fig:BendingAndBuckling}~a).
Applying an electric field $E$ to the piezo rings leads to an induced strain and hence to a change of their diameter~$D$ by 
\begin{equation}
\label{eq:diameter}
\Delta D \,=\,d_{31} D E .
\end{equation}
The piezoelectric coefficient $d_{31}$ depends on the used piezo material and is also a function of the applied electric field \cite{BinalSMS}. If one applies opposite voltages to the upper and lower piezo ring ($E_{\text{up}}=-E_{\text{low}}$), which leads to a expansion of one and a contraction of the other ring, the membrane deforms approximately spherically, as shown in \fref{fig:BendingAndBuckling}~b) as it is bent by the piezo rings at its outer boundary. In contrast, a contraction of both rings forces the glass membrane to buckle out of the plane, leading to a more hyperbolic shape (\fref{fig:BendingAndBuckling}~c)).

\begin{figure*}[bhpt!]
	\centering
	\def\svgwidth{0.98\columnwidth}
	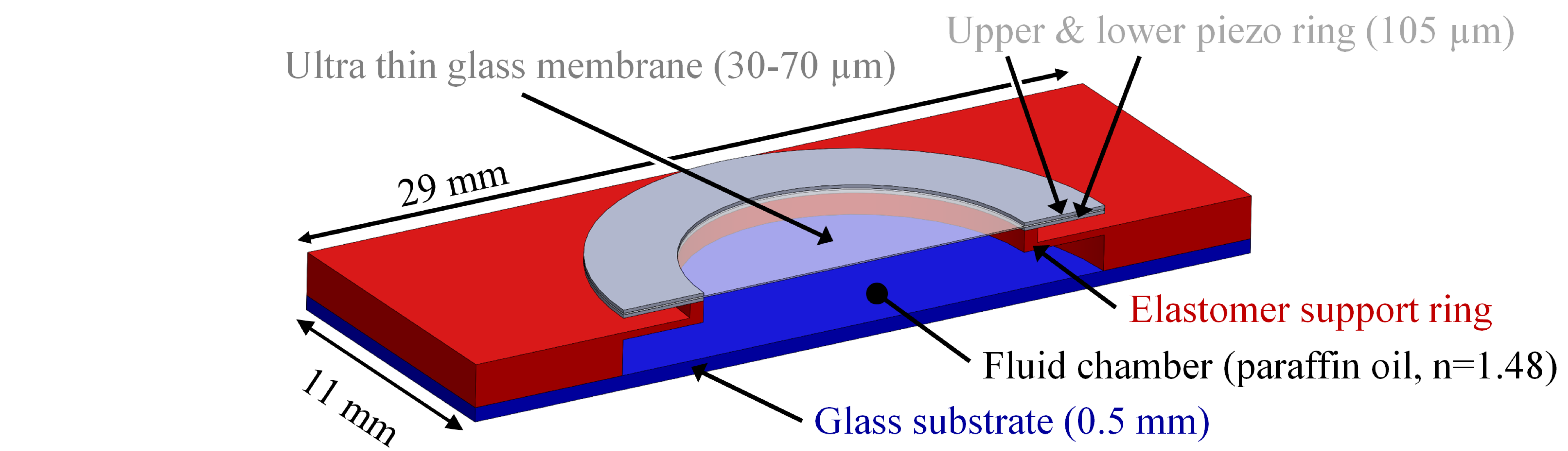
	\caption{Cross-section through the lens model (to scale).}
	\label{fig:Lens_total}
\end{figure*}

To generate a lens effect, we combine the active piezo-glass-sandwich with an elastic fluid chamber, add a transparent fluid (paraffin oil, $n=1.48$) and seal the chamber with a glass substrate. The complete assambly is shown in \fref{fig:Lens_total}.
 \\
\myEmphTwo{For a simplified analytical model of the bending mode, as derived in~\cite{Matthias14}, we assume a spherical deformation and neglect forces, e.g. the bending stiffness of the piezo rings and the glass membrane. To obtain the radius of curvature, we need to obtain the radius $R$ of a segment with arc length $D$ that describes a cut through the center of the lens. Considering one piezo ring with diameter $D+\Delta D$ at a distance $s/2$ above the center of the membrane and one with diameter $D-\Delta D$ by $s/2$ below the center of the membrane, and inserting \eref{eq:diameter} with $E_{\text{up}}=-E_{\text{low}}$, we find 
\begin{equation}
\label{eq:bendingCurvPre}
\frac{D+d_{31} D E_{\text{up}}}{R+\frac{1}{2}s}=\frac{D+d_{31} D E_{\text{low}} }{R-\frac{1}{2}s}\ .
\end{equation}
Here, s is the distance of the neutral planes of the piezo sheets (see \fref{fig:BendingAndBuckling}), roughly of the order of the membrane thickness plus one piezo thickness. 
For small angles, where $R \gg s$, we can approximate the curvature in the bending mode by }
\begin{equation}
\label{eq:bendingCurv}
R^{-1}\approx\, s^{-1} \; d_{31} \; (E_{\text{up}}-E_{\text{low}})\ .
\end{equation}
In the (pure) buckling mode,\myEmphTwo{ where $E_{up}=E_{low}=E$, we assume a membrane with diameter $D_{\text{in}}$ that is compressed at its circumference according to~\eref{eq:diameter} and then buckles into a spherical shape. Considering again a circular arc that cuts through the lens, we find
\begin{equation}
\label{eq:bucklingCurvPre}
(D_{\text{in}}-\Delta D_{\text{in}})=2R \sin\frac{\alpha}{2} \text{ , where }\alpha=\frac{D_{\text{in}}}{R} \ . 
\end{equation}
To leading order in $\Delta D$, this gives us}
\begin{equation}
\label{eq:bucklingCurv}
\myEmphTwo{R^{-1}\approx \, D_{\text{in}}^{-1} \sqrt{24 \; d_{31} \; E}\ .} 
\end{equation}
As a result, using the lensmaker’s equation for thin lenses, the focal power for the bending mode is
\begin{equation} \label{eq:bend}
f^{-1}=\,\Delta n \; R^{-1}\approx \Delta n \; s^{-1} \; d_{31} \; (E_{\text{up}}-E_{\text{low}})\ .
\end{equation}
Similarly, the focal power in the buckling mode is
\begin{equation} \label{eq:buck}
\myEmphTwo{f^{-1} \approx\, D_{\text{in}}^{-1} \sqrt{24 \; d_{31} \; E}\ .}
\end{equation}
The direction of the buckling can be chosen by first bending the membrane and then buckling it. All of these estimates only take into account the geometric changes and do not consider the forces in the piezo rings and the glass membrane. \myEmph{In particular, for both modes, we considered only the von Neumann boundary condition (bending) or Dirichlet boundary condition (buckling). Considering the Dirichlet boundary condition also in the bending mode, i.e., also the radial in-plane strain will result in a more elliptic profile. Similarly considering the von Neumann boundary condition in the buckling -- the resistance of the piezo rings to bending -- will result in a more hyperbolic deformation. In both cases, the estimated focal power will hence not be accurate, but we can on the one hand estimate the order of magnitude and on the other hand, we see the scaling behavior: The bending is linear and the buckling behaves as a square root of the applied electric field. Furthermore, the bending focal power scales inversely with the overall thickness and is independent of the aperture whereas the buckling is, to first order, independent of the thickness and scales inversely to the aperture.}

Including forces in sufficient detail would be hard to impossible on an analytic level. This is why we need reliable simulations to predict and optimize the behavior of the lens. \myEmph{Nevertheless, there are examples in literature, where the bending deflection for similar actuators used for micro pumps has been calculated~\cite{Pump1,Pump2}. The calculations of these actuators with only one piezo and one passive layer are based on the bending moment that is induced in the structure by the expansion of the piezo. However, non-linearity, which occurs in the bending and buckling mode of our two piezo layer structure, is not taken into consideration. By comparing the results of the following simulations, we found that a neglection of this non-linearity leads to an increase of up to $+\SI{73}{\percent}$ for the radius of curvature.}

\section{Simulation} \label{sec:sim}
We used \textit{COMSOL Multiphysics} (Version 5.3a) with the multiphysics module ``Piezoelectric effect'' that combines the ``Solid Mechanics'' and the ``Electrostatics'' module to simulate the piezo electric deformation. The piezo electric material properties were set to the strain-charge form. 
\subsection{Lens model}\label{sec:lensmodel}
\begin{figure*}[bhpt!]
	\centering
	\def\svgwidth{0.8\columnwidth}
	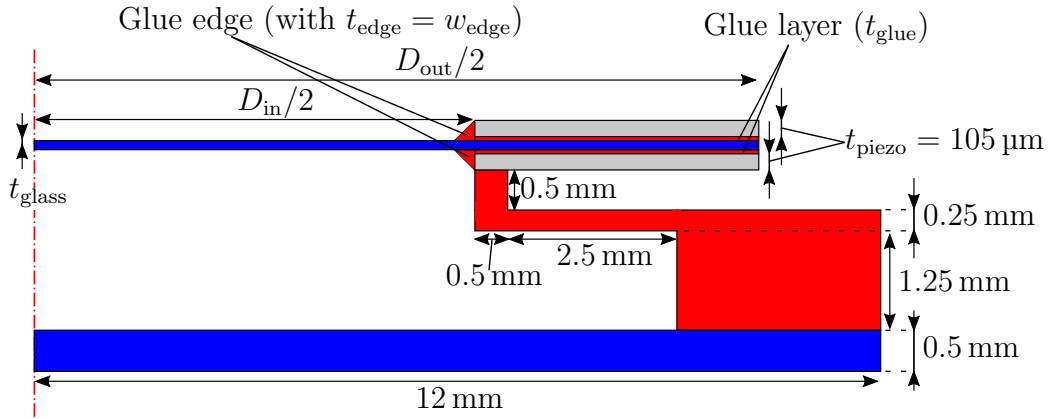
	\caption{Half section of the model geometry with fully defined geometry parameters (rotated around the vertical \textit{dashed red line}).}
	\label{fig:model_dimensions}
\end{figure*}

\begin{figure*}[bhpt!]
	\centering
	\def\svgwidth{0.85\columnwidth}
\begingroup%
  \makeatletter%
  \providecommand\color[2][]{%
    \errmessage{(Inkscape) Color is used for the text in Inkscape, but the package 'color.sty' is not loaded}%
    \renewcommand\color[2][]{}%
  }%
  \providecommand\transparent[1]{%
    \errmessage{(Inkscape) Transparency is used (non-zero) for the text in Inkscape, but the package 'transparent.sty' is not loaded}%
    \renewcommand\transparent[1]{}%
  }%
  \providecommand\rotatebox[2]{#2}%
  \newcommand*\fsize{\dimexpr\f@size pt\relax}%
  \newcommand*\lineheight[1]{\fontsize{\fsize}{#1\fsize}\selectfont}%
  \ifx\svgwidth\undefined%
    \setlength{\unitlength}{2737.5bp}%
    \ifx\svgscale\undefined%
      \relax%
    \else%
      \setlength{\unitlength}{\unitlength * \real{\svgscale}}%
    \fi%
  \else%
    \setlength{\unitlength}{\svgwidth}%
  \fi%
  \global\let\svgwidth\undefined%
  \global\let\svgscale\undefined%
  \makeatother%
  \begin{picture}(1,0.2739726)%
    \lineheight{1}%
    \setlength\tabcolsep{0pt}%
    \put(0,0){\includegraphics[width=\unitlength,page=1]{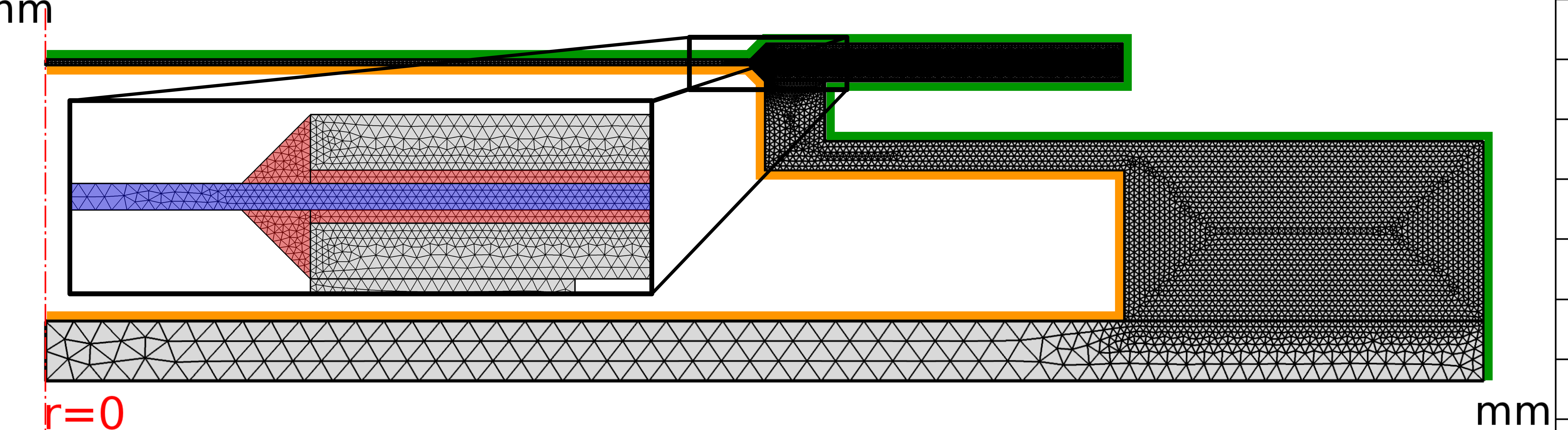}}%
    \put(0.7805992,0.23408511){\color[rgb]{0,0,0}\makebox(0,0)[lt]{\smash{\begin{tabular}[t]{l}Enclosed\\ volume\end{tabular}}}}%
    \put(0,0){\includegraphics[width=\unitlength,page=2]{Mesh4_with_gluelayer_and_glueedge_color_new.pdf}}%
    \put(0.52070777,-0.01255682){\color[rgb]{0,0,0}\makebox(0,0)[lt]{\smash{\begin{tabular}[t]{l}Fixed\end{tabular}}}}%
    \put(0.47539742,0.11836284){\color[rgb]{0,0,0}\makebox(0,0)[lt]{\smash{\begin{tabular}[t]{l}\color[rgb]{1,0.588,0}{Normal surface}\\\color[rgb]{1,0.588,0}{load (pressure)}\end{tabular}}}}%
    \put(0,0){\includegraphics[width=\unitlength,page=3]{Mesh4_with_gluelayer_and_glueedge_color_new.pdf}}%
    \put(0.13724886,0.25096315){\color[rgb]{0,0,0}\makebox(0,0)[lt]{\smash{\begin{tabular}[t]{l}\color[rgb]{0,0.588,0}{Free}\end{tabular}}}}%
    \put(0,0){\includegraphics[width=\unitlength,page=4]{Mesh4_with_gluelayer_and_glueedge_color_new.pdf}}%
  \end{picture}%
\endgroup%

	\caption{Example of the mesh and illustration of the boundary conditions. Inset: Detailed view of glue layer, glue edge and fluid chamber (polyurethane, \textit{red}), of piezo rings (\textit{gray}) and of the glass membrane (\textit{blue}).}
	\label{fig:Mesh}
\end{figure*}

We defined the geometry according to \fref{fig:model_dimensions} with the following set of free parameters:
\begin{itemize}
\item Upper and lower piezo ring with inner and  outer radii~$D_{\text{in}}/2$ and $D_{\text{out}}/2$ and thickness~$t_{\text{piezo}}=\SI{105}{\micro\meter}$
\item Ultra-thin glass membrane with radius $D_{\text{out}}/2$ and thickness~$t_{\text{glass}}$
\item Glue layer between glass membrane and piezo with thickness $t_{\text{glue}}$
\item Glue edge with $t_{\text{edge}}$ and $w_{\text{edge}}$ at the corner, where the piezo rings are in contact with the inner part of the glass membrane
\end{itemize}
\normalsize
The other dimensions of the elastic polyurethane fluid chamber that we keep constant are also shown in \fref{fig:model_dimensions}. We defined all mesh sizes \myEmphTwo{to be smaller than half of their corresponding minimal thickness} to ensure at least two layers of mesh elements in any component of the lens. An example of the mesh is shown in \fref{fig:Mesh}.
\subsection{Piezoelectric non-linearity}\label{subsec:nonlin}
\begin{figure*}[hbpt!]
	\centering
	\includegraphics[width=1\linewidth]{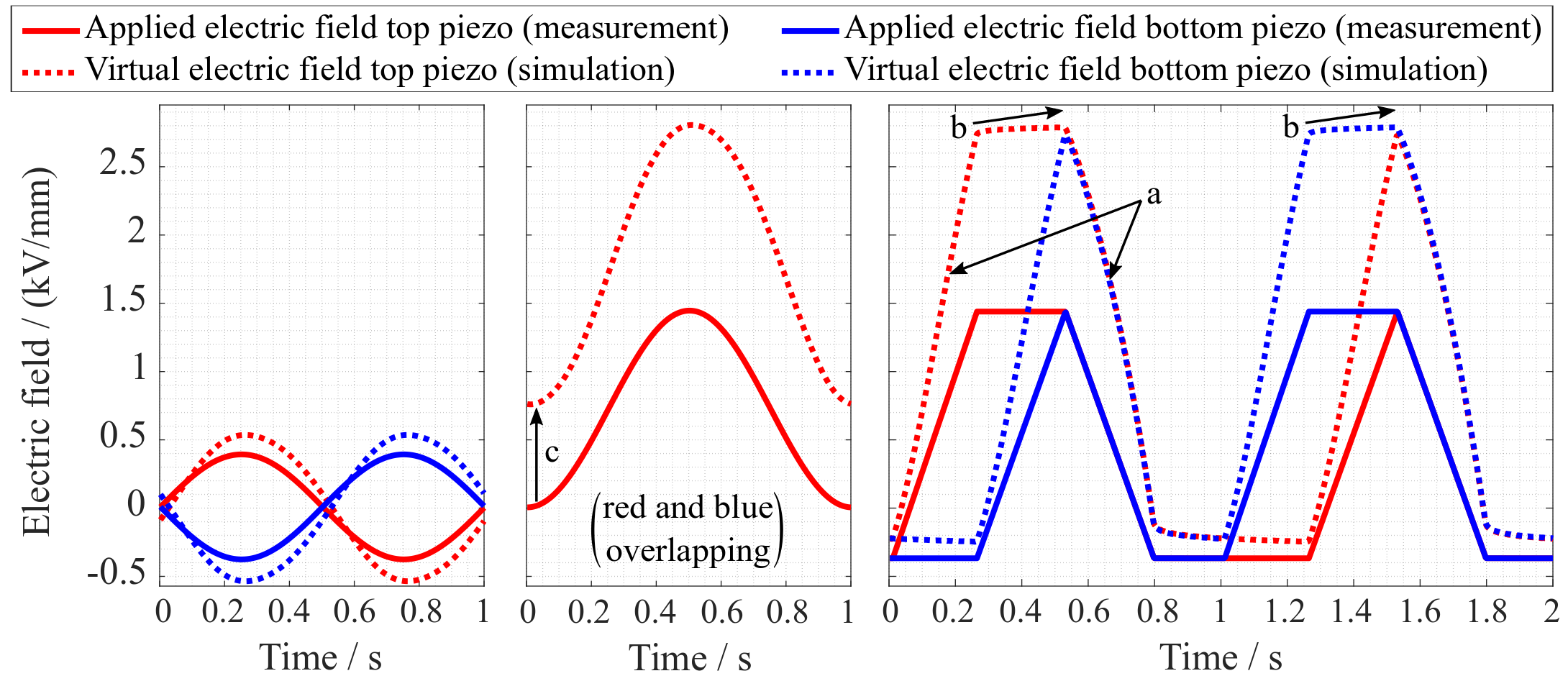}
	\caption{Electric field applied in the measurement (\emph{solid line}) and virtual (corrected) electric field applied in the simulation (\textit{dashed line}) for the bending mode (left), the buckling mode (middle) and the working range trajectory (right).}
	\label{fig:Voltage_real_vs_sim}
\end{figure*}

Because of the non-linearity of the piezoelectric charge coefficient $d_{31}$ and the hysteresis and creep effects, we did not simply use a constant $d_{31}$, but we needed to determine the charge coefficient as a function of the applied voltage and its voltage history. For this reason,  we followed the established strategy of \cite{BinalSMS} and manufactured a simple mono-morph bending beam with one $\SI{100}{\micro\meter}$ thick passive glass layer and one \textit{Ekulit} piezo layer that we could simulate reliably. We applied the same electric fields as for the lens measurements (see section~\ref{sec:meas}) to this beam, measured the beam curvature and compared it to the simulation as done in~\cite{BinalSMS}. That way, we found a modified electric field sequence shown in \fref{fig:Voltage_real_vs_sim} that is needed as an input for the simulation to achieve the same curvature as the measured beam while using the standard coupling matrix with $d_{31} =\SI{274}{\pico\meter\over\volt}$ for PZT-5H \myEmph{from the material library of \textit{COMSOL Multiphysics}}. This virtual electric field sequence now contains the non-linearity of the charge coefficient in addition to all hysteresis and creeping effects. \myEmph{In this way, we did not need to scale the coupling matrix for our piezo material (non--linear, $d_{31} = \SI{487}{\pico\meter\over\volt}$, see section~\ref{sec:fab}), because the virtual voltage already contains any scaling of $d_{31}$. More importantly, we also bypassed the need to implement $d_{31}$ as a function of the electric field and furthermore to implement the unknown and complex hysteresis behavior that is not natively implemented in \textit{COMSOL Multiphysics}.} 

In \fref{fig:Voltage_real_vs_sim} we find a hysteresis (\textit{a}) from the asymmetry of the rising and falling slopes and short term creep (\textit{b}) from the time-dependence when the physical voltage is fixed. We further find a voltage shift to positive values for the non-symmetric cycles (buckling, trajectory) which is most likely due to long-term creep (\textit{c}).
\subsection{Material properties and fabrication effects} \label{subsec:fabricationeff}
The elastomer chamber is made from polyurethane \textit{Smooth-on ClearFlex 50} (CF50) with a Young's modulus of \SI{2.22}{\mega\pascal} that we determined in a pull test at approximately 5\% strain. We took the Young's modulus of the \textit{Schott D263t} glass membrane  $Y_{\text{glass}}\approx\SI{72.9}{\giga\pascal}$~\cite{Schott} and the \textit{Smooth-on CrystalClear 200} (CC200) glue layer $Y_{\text{CC200}}\approx\SI{1.38}{\giga\pascal}$ from their material data sheets~\cite{CC200}. \myEmphTwo{Again, we used the standard materials from the \emph{COMSOL} material library for polyurethane and glass and replaced the Young's moduli by the mentioned values.} In~\cite{BinalSMS}, we determined the Young's modulus of the \textit{Ekulit} PZT to $Y_{\text{piezo}}\approx\SI{37}{\giga\pascal}$, \myEmphTwo{which we implemented by scaling the compliance matrix of the PZT-5H}. 

We measured piezo thicknesses between \SI{100}{} and \SI{110}{\micro\meter} resulting in a mean value of $t_{piezo}=\SI{105}{\micro\meter}$. Similarly, the glue layers varied from \SI{18}{} to \SI{32}{\micro\meter}, so we used $t_{glue}=\SI{25}{\micro\meter}$. We additionally estimated a glue edge with $t_{\text{edge}}=w_{\text{edge}}=\SI{130}{\micro\meter}$, which equals $t_{\text{glue}}+t_{\text{piezo}}$ resulting in a isosceles triangle shape from the edge of the piezo. Besides that, we took into account a thermally induced pre-strain in the glass-piezo sandwich which results from the gluing process in the oven at a temperature of \SI{50}{\celsius} and the laboratory room temperature of $24\pm\SI{3}{\celsius}$.  This temperature difference of \SI{26}{\celsius} leads to an internal stress in the membrane composite that effects its deflection. Again, we took the thermal expansion coefficient ($\alpha_{glass}=\SI{7.2e-6}{\per\kelvin}$) from the material data sheet~\cite{Schott}. The thermal expansion of the piezo is not given by the manufacturer, but similar PZT materials have coefficients in the range of \SI{4}{} to \SI{8e-6}{\per\kelvin}~\cite{PI}. We added this thermal initial strain to the piezo rings and to the glass membrane in the simulation \myEmphTwo{with the in-built ``initial stress and strain'' functionality of \emph{COMSOL}}. After variation of the thermal expansion coefficient of the piezos in the simulation we found that the mean of the mentioned value range of \SI{6e-6}{\per\kelvin} fitted best to our experiments and was used further on. The effect of these parameters will also be discussed in detail in section \ref{sec:eval}.
\subsection{Simulation strategy}
\myEmphTwo{For the simulation, we used parametric sweeps of stationary studies, in which applied a sequence of voltage combinations for the upper and lower piezo (as shown in \fref{fig:Voltage_real_vs_sim}). Because of the non-linearity and bistability, we always took the solution of the previous voltage step as a starting point for the simulation to ensure the correct direction of buckling. For the initial simulation step of the bending mode and the trajectory, we used the non-deflected lens and for the initial step of the buckling mode, we used the maximum deflection of the bending mode to determine the direction of buckling. Without this pre-deflection, the simulation does not show any buckling as an un-deflected ideal lens is in a metastable state that, however, is always distorted in a real prototype. Furthermore,} we activated the ``geometric non-linearity'' as the buckling is an intrinsically non-linear effect \myEmph{with high in-plane tensions that requires this kind of non-linear simulation.}   
After verifying that the 2-dimensional rotationally symmetric model agrees with a full 3-dimensional model, we chose the former one as it reduces the computation time dramatically. As we have an enclosed volume inside the lens, \myEmphTwo{we need to keep the volume in the fluid chamber constant in the simulation. We determine the volume change due to the deflection by integrating over the displacement of the membrane. We further use the ODE module to apply an internal counter pressure to all inner boundaries of the chamber as a control variable (compare \fref{fig:Mesh}) to set the determined volume change to 0, keeping the initial volume in the fluid chamber constant.}

\section{Fabrication}\label{sec:fab}
As a piezo material, we detach the piezo sheets from \textit{Ekulit} sound buzzers  with the mentioned small thickness of $ t = \SI{105}{\micro\meter}$ to achieve high focal powers referring to equation (4). They in fact showed better performance with a high piezoelectric coefficient of \myEmph{$d_{31} \approx \SI{-487}{\pico\meter\over\volt}$~\cite{BinalSMS} for high electric fields and a better surface quality than many readily available raw piezo foils. We used this value for all analytical calculations but need to be aware that for small electric fields, the piezoelectric coefficient is reduced by up to \SI{40}{\percent}~\cite{BinalSMS} resulting in overestimated focal powers in the bending mode.} We structure the piezo rings and the \SI{30}{}, \SI{50}{} and \SI{70}{\micro\meter} thick glass membrane with an UV laser. At the edges of the piezo we additionally remove \SI{100}{\micro\meter} of the electrode to avoid electric breakdown at high electric fields. To remove residues caused by the laser cutting process, we clean the piezo rings subsequently in an ultrasonic bath for \SI{5}{\second}  in \SI{25}{\percent} HNO$_3$. After structuring, we glue the glass in between the piezo rings using hard polyurethane (CC200) and a system of vacuum chucks with alignment structures in an oven at \SI{50}{\celsius}. When cured, we glue this glass-piezo-sandwich to the elastic polyurethane fluid chamber (CF50), that we cast from a mold. We seal this chamber with the glass substrate using the same soft polyurethane as glue. 

\begin{figure*}[bhpt!]
	\centering
	\includegraphics[width=0.9\linewidth]{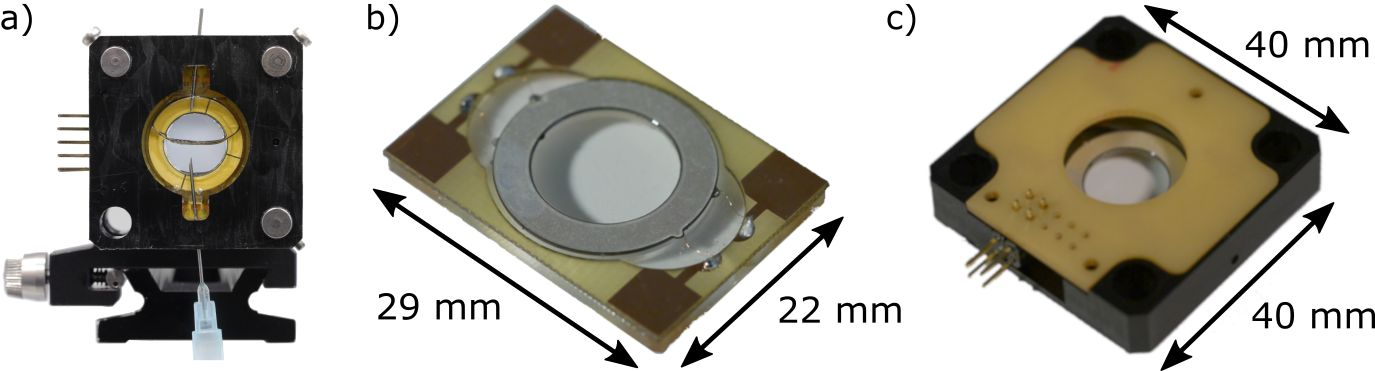}
	\caption{(a)~Filling process of the lens, (b)~close up of the fabricated lens and (c)~packaged lens for the standard \SI{30}{\milli\meter} cage system with electric contacts.}
	\label{fig:fabrication}
\end{figure*}

In this paper, we do not fill the fluid chamber with an optical oil (usually paraffin oil, n = 1.48), because we are interested in the purely mechanical deformation of the membrane. For this reason, all focal power calculations are based on a “virtually filled” lens, where we take into account the curvature of the lens and calculate the refractive power with an assumed refractive index of n = 1.48. 
The internal pressures that we found in the simulations were of the order of \SI{20}{\pascal} for maximum deflections. \myEmph{Neglecting this internal pressure and assuming a fluid chamber that is open at the backside, the curvature of the membrane changes by approximately \SI{3}{\percent}. As a result, we need to take the fluid into account and consider its effects on the membrane deformation. Nevertheless, the compressibility of an ideal gas compared to an incompressible fluid in the chamber results in changes of less then $0.3\%$, so we can consider air and fluid to be approximately equivalent. Hence, all the results that we obtained from lenses filled with air will be equivalent to lenses filled with the optical fluid.
This filling is normally done after the gluing process} by inserting two syringes through the elastic material of the fluid chamber along two fluidically optimized channels designed for bubble-free filling. One syringe supplies the optical oil, the other one releases the air from the fluid chamber as shown in \fref{fig:fabrication}~a). Finally, we seal the holes by adding a drop of polyurethane. The finished lens (\fref{fig:fabrication}~b)) is then packaged in a custom mount for the standard \SI{30}{\milli\meter} cage system, shown in \fref{fig:fabrication}~c).

\normalsize
\section{Characterization of focal power and aspherical behavior}\label{sec:meas}
We characterized the membrane deformation by measuring the surface with a chromatic confocal pen (\textit{Polytec, TopSens CL4/MG35}) with a vertical accuracy of approximately $\SI{0.1}{\micro\meter}$ and a spot size of \SI{12.3}{\micro\meter} and with a high-precision translation stage. To obtain the time-dependent surface profile, we preformed a pointwise measurement during a periodic actuation with a sensor frequency of \SI{400}{\hertz}.

We limited the electric field against the polarization to $\SI{-0.38}{\kilo\volt\over\milli\meter}$, i.e., less than half of the negative coercive field strength of $\SI{0.78}{\kilo\volt\over\milli\meter}$~\cite{BinalSMS}. We similarly set a maximum electric field of $\SI{1.43}{\kilo\volt\over\milli\meter}$ to avoid electrostatic breakdown.

We preformed three different measurements that we will compare to the simulation results later. First, we applied opposite voltages for the upper and lower piezo ring to drive the lens in a pure bending mode with an amplitude of $\SI{0.38}{\kilo\volt\over\milli\meter}$ with a symmetric sinusoidal wave with \SI{1}{\hertz} as a quasi static actuation. For the pure buckling mode, where we needed a simultaneous contraction of both rings, we applied a sinusoidal wave with \SI{1}{\hertz} between $\SI{0}{}$ and $\SI{1.43}{\kilo\volt\over\milli\meter}$. In this case, however, we waited 30 min before starting a measurement with an overall measurement time of 120 min to avoid longtime creeping effects of the piezo.

To determine the full working range of the lens, we applied a voltage trajectory, where we cover several points of interest and outline the stable operating region. First, we pulled the membrane flat by expanding both piezo rings at $-\SI{0.38}{\kilo\volt\over\milli\meter}$. Then, the voltage of the upper piezo was increased linearly to the maximum voltage ($\SI{1.43}{\kilo\volt\over\milli\meter}$) to reach the maximum bending effect. For the maximum buckling in the next step, we increased the electric field of the lower piezo ring to the same value. Finally we lowered both electric fields to $-\SI{0.38}{\kilo\volt\over\milli\meter}$, pulling the membrane flat again. For symmetry reasons we exchanged the upper and lower voltages and applied the cycle a second time. In principle, it is also possible to operate the lens in a metastable mode, where we bend it against the direction of the deformation in a buckled state. However for reasons of reliability we do not consider such modes in this paper. The entire trajectory takes \SI{2}{\second} and we again waited 30 min to start the measurement after starting the actuation. The voltages for all three measurement modes are shown in section~\ref{subsec:nonlin}.

The measurement data was then evaluated with a $4^{th}$ order rotationally symmetric fit:
\begin{eqnarray} 
 z(x,y) = & \alpha_0 +\alpha_{1,x} x+ \alpha_{1,y}  y+\alpha_{2}  r^2+\alpha_{4}  r^4\; , \label{eq:oberflaechen_formel_linse}
\end{eqnarray} where $r=\left(\sqrt{x^2+y^2}\right)$ with the lens center at $x=0$ and $y=0$.
Using the lensmaker's equation, the focal power is then
\begin{equation}\label{eq:focalPower}
f^{-1}\approx 2\;\Delta n\;\alpha_{2}\\ .
\end{equation}
Similarly, a measure for the aspherical behavior is given by the parameter $\alpha_{4}$. A conversion to Zernike polynomials can be realized by a straightforward linear transformation\myEmph{, where the spherical Zernike coefficient of the optical path length is given by
\begin{equation}\label{eq:zernike}
Z_4^0= \frac{\Delta n\;r_{\text{max}}^4}{6} \; \alpha_{4}  , 
\end{equation}
where $r_{\text{max}}$ is the evaluated aperture radius.
}

\section{Results and discussion}\label{sec:eval}
\subsection{Evaluation of measurement and simulation}
To verify the simulation and to study the influence of geometric parameters, we fabricated a set of different lens designs, systematically varying the inner and outer diameter of the piezo rings and the glass membrane thickness in comparison to our standard design with $D_{\text{out}}= \SI{18}{\milli\meter}$, $D_{\text{in}}= \SI{12}{\milli\meter}$ and $t_{\text{glass}}= \SI{50}{\micro\meter}$. \Tref{tab:overview} gives an overview over the different lenses. To demonstrate the reproducibility of the manufacturing process, we fabricated two lenses in the standard design 1 (a, b). We evaluated the surface profile only over a diameter of \SI{10.4}{\milli\meter} to avoid edge effects caused by glue that has flown onto the membrane as mentioned in section \ref{sec:lensmodel}. For the lens with the inner diameter of \SI{10}{\milli\meter}, we had to reduce the evaluated diameter slightly to \SI{9.8}{\milli\meter}. In this case, there was only a small amount of glue causing a less then \SI{100}{\micro\meter} glue edge, while for other lenses the glue edge was up to \SI{400}{\micro\meter} wide.
\begin{table}[h!]
\centering
\small
\caption{Overview over the measured lenses: The \textbf{highlighted} dimensions show the changed parameter and the \textbf{color} represents the corresponding color of the result graphs.}
\label{tab:overview}
\begin{tabular}{|l|l|l|l|l|l|}\hline
Lens design& Outer piezo diameter &Inner piezo diameter &glass membrane thickness \\
 &$D_{\text{out}}$ / \SI{}{\milli\meter}       &$D_{\text{in}}$ / \SI{}{\milli\meter} & $t_{\text{glass}}$ / \SI{}{\micro\meter}\\
\hline
\textbf{1} (a)     & 18                & 12              & 50  			\\
\color[rgb]{0,0,1}\textbf{1} (b)     & 18                & 12              & 50  			\\
\hline
\color[rgb]{1,0.8,0} \bf2     & \textit{18}                & \textit{12}              & \textit{\textbf{30}} \\
\color[rgb]{0,0.58,0}\bf3     & 18                & 12              & \textbf{70} \\
\hline
\color[rgb]{1,0,0}\bf4     & 18                & \textbf{10}       & 50  			\\
\color[rgb]{0,1,1}\bf5     & 18                & \textbf{14}       & 50  			\\
\hline
\color[rgb]{1,0.,1}\bf6     & \textbf{15}         & 12              & 50  			\\
\color[rgb]{0.53,0.33,0.22}\bf7			& \textbf{21}         & 12              & 50  			\\
\hline
\end{tabular}
\end{table}

\begin{figure*}[bhpt!]
	\centering
	\includegraphics[width=0.49\linewidth]{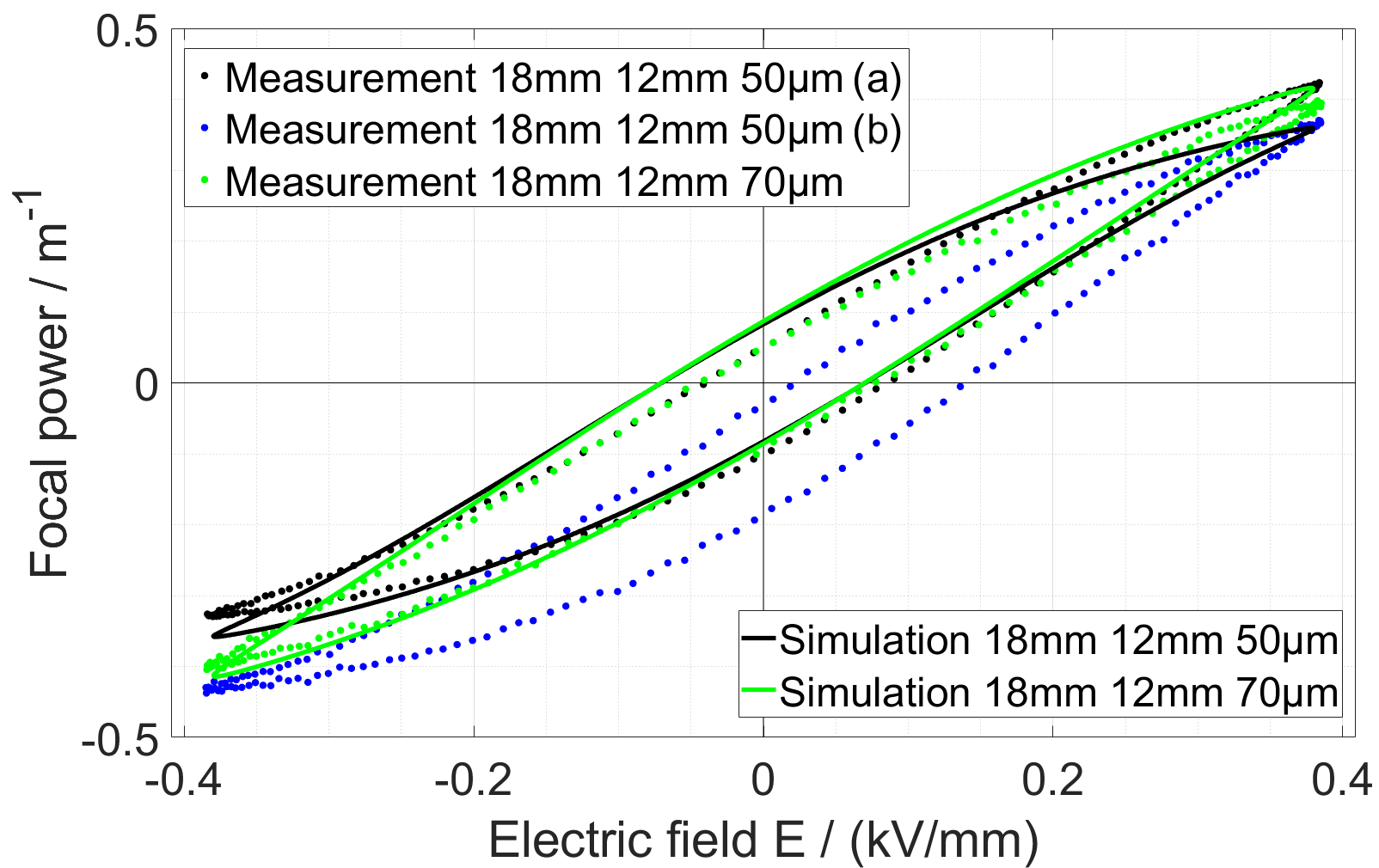}
	\includegraphics[width=0.49\linewidth]{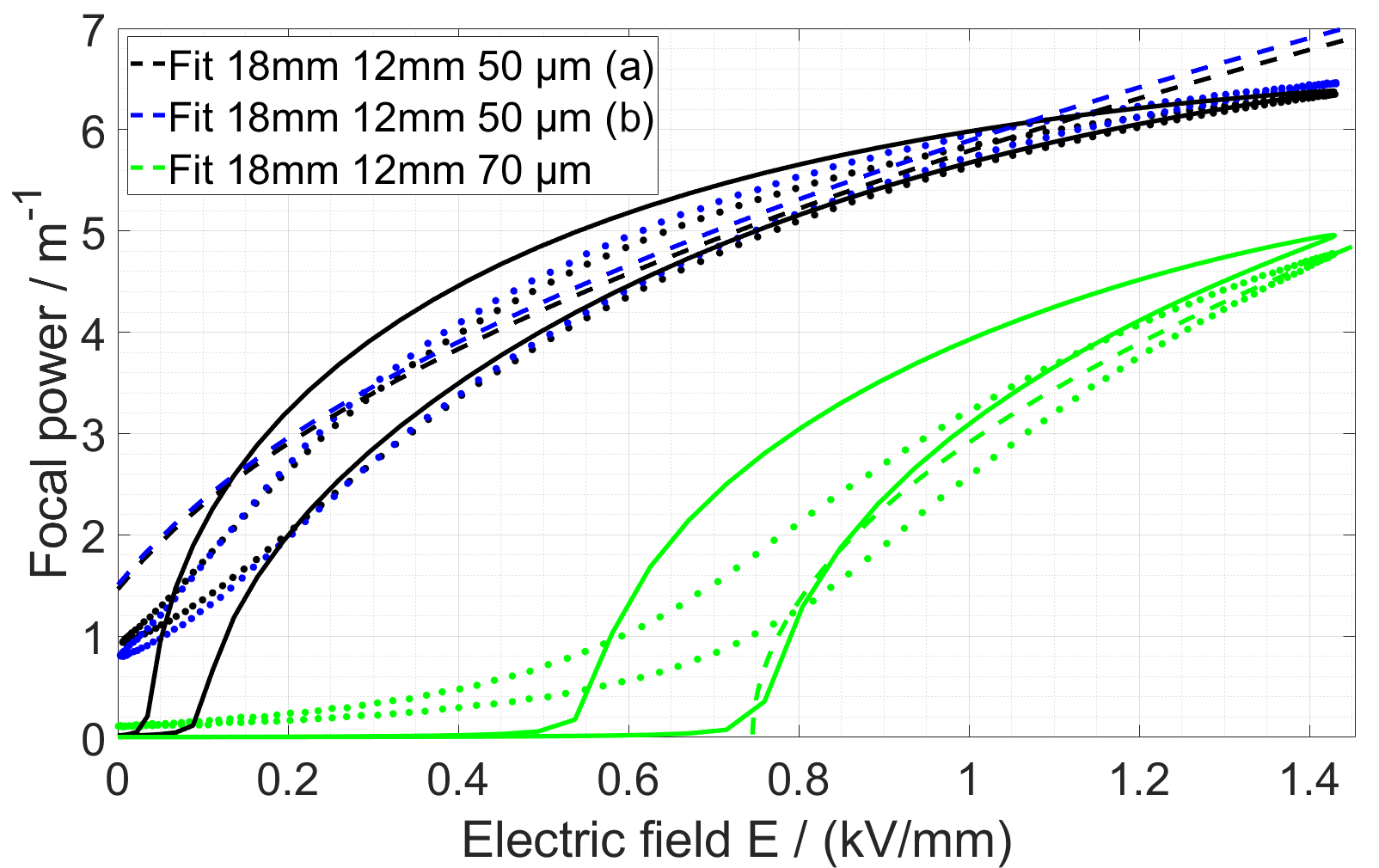}
	
	\caption{Simulated (\textit{solid}) and measured (\textit{dotted}) focal power as a function of the applied electric field (top electrode) for the pure bending mode (left) and the pure buckling mode (right): Lenses with $D_{\text{out}}=\SI{18}{\milli\meter}$, $D_{\text{in}}=\SI{12}{\milli\meter}$ and $t_{\text{glass}}=\SI{50}{\micro\meter}$ (\textit{black}, \textit{blue}) and $t_{\text{glass}}=\SI{70}{\micro\meter}$ (\textit{green}). Fit according to analytical model shown as dashed line.}
	\label{fig:Sim_Bend_Buck}
\end{figure*}

In the left graph of \fref{fig:Sim_Bend_Buck} we show the focal power as a function of the electric field~$E$ applied to the upper piezo ring for a pure bending deflection for the lenses~1 and~3. We clearly see an approximately linear behavior as predicted by \eref{eq:bend}. A suitable fit with the function $f^{-1}=a\,E+c$  results in \myEmphThree{slopes of \SI{1.02e-6}{} (black, (a)) and \SI{1.09e-6}{\per\volt} (blue, (b)) for the \SI{50}{\micro\meter} membrane and \SI{1.06e-6}{\per\volt} (green) for the \SI{70}{\micro\meter} membrane. The deviation compared to the analytical prediction of \SI{2.28e-6}{} and \SI{2.08e-6}{\per\volt} for \SI{50} and \SI{70}{\micro\meter}}, respectively, is a factor of approximately 2.2, if we assume a distance between the neutral planes of \SI{205}{\micro\meter} for the  \SI{50}{\micro\meter} thick glass and \SI{225}{\micro\meter} for the \SI{70}{\micro\meter} thick glass ($t_{\text{glass}}+2\,t_{\text{glue}}+t_{\text{piezo}}$). The fit results for these and all the following lenses are shown in \tref{tab:coeffs}. 
This calculation neglects forces and only takes into consideration the deformation generated by the piezo rings. Up to a small negative pre-deflection, the simulation reproduces the hysteresis behavior very well for the \SI{70}{\micro\meter} membrane lens (green) and has small deviations for the first \SI{50}{\micro\meter} membrane lens (black, (a)), most likely caused by the pre-deflection of the prototype and variations of the glue layer thickness and temperature while manufacturing. The second \SI{50}{\micro\meter} membrane lens (blue, (b)) shows a lager negative pre-deflection but a similar absolute focal power range.
Looking at the pure buckling mode in the right graph of \fref{fig:Sim_Bend_Buck}, we find an approximate square-root shape dependency, which was predicted by \eref{eq:buck} for the measurement and the simulation. \myEmph{The simulation of the \SI{50}{\micro\meter} membrane lenses agrees in a wide range with the measurement. Comparing the two \SI{50}{\micro\meter} lenses, we find reproducible results up to some offset in the pre-deflection.
For the \SI{70}{\micro\meter} measurement, we find a small shift in the offset voltage compared to the simulation, which corresponds to an internal stress, e.g.~due to a deviation in the curing temperature. Furthermore, we find a smooth transition from a non-buckled state to the buckled state in the measurement whereas the simulation gives a much sharper onset of buckling. The reason may be that the perfectly flat and symmetric membrane in the simulation needs a larger amount of energy to accumulate until it becomes unstable and buckles, whereas the fabricated membrane always has some pre-deflection, imperfections and asymmetry such that it buckles more easily. This onset of the buckling occurs} between 0 and \SI{0.1}{\kilo\volt\over\milli\meter} for the \SI{50}{\micro\meter} membrane and between \SI{0.5}{} and \SI{0.8}{\kilo\volt\over\milli\meter} for the \SI{70}{\micro\meter} membrane. The dashed line shows the fit of the function $f^{-1}=b\,\sqrt{E-E_0}$ comparing the measurement to the analytic evaluation in \eref{eq:buck} and introducing an offset electric field $E_0$ as a starting point for the buckling. The fit was evaluated only over the central third of the measurement to neglect the transition zone and the zone where high forces dominate. The results given in \tref{tab:coeffs} match again within a factor of approximately 2. The significant difference in the offset electric field results from the larger buckling threshold load in the thicker membrane. The slightly negative $E_0$ for the thin membrane results from the combination of thermal pre-stress, the long-term creep in the piezo (see \fref{fig:Voltage_real_vs_sim}) and the critical buckling load of the membrane.

\begin{figure*}[t!]
	\centering
	\includegraphics[width=0.49\linewidth]{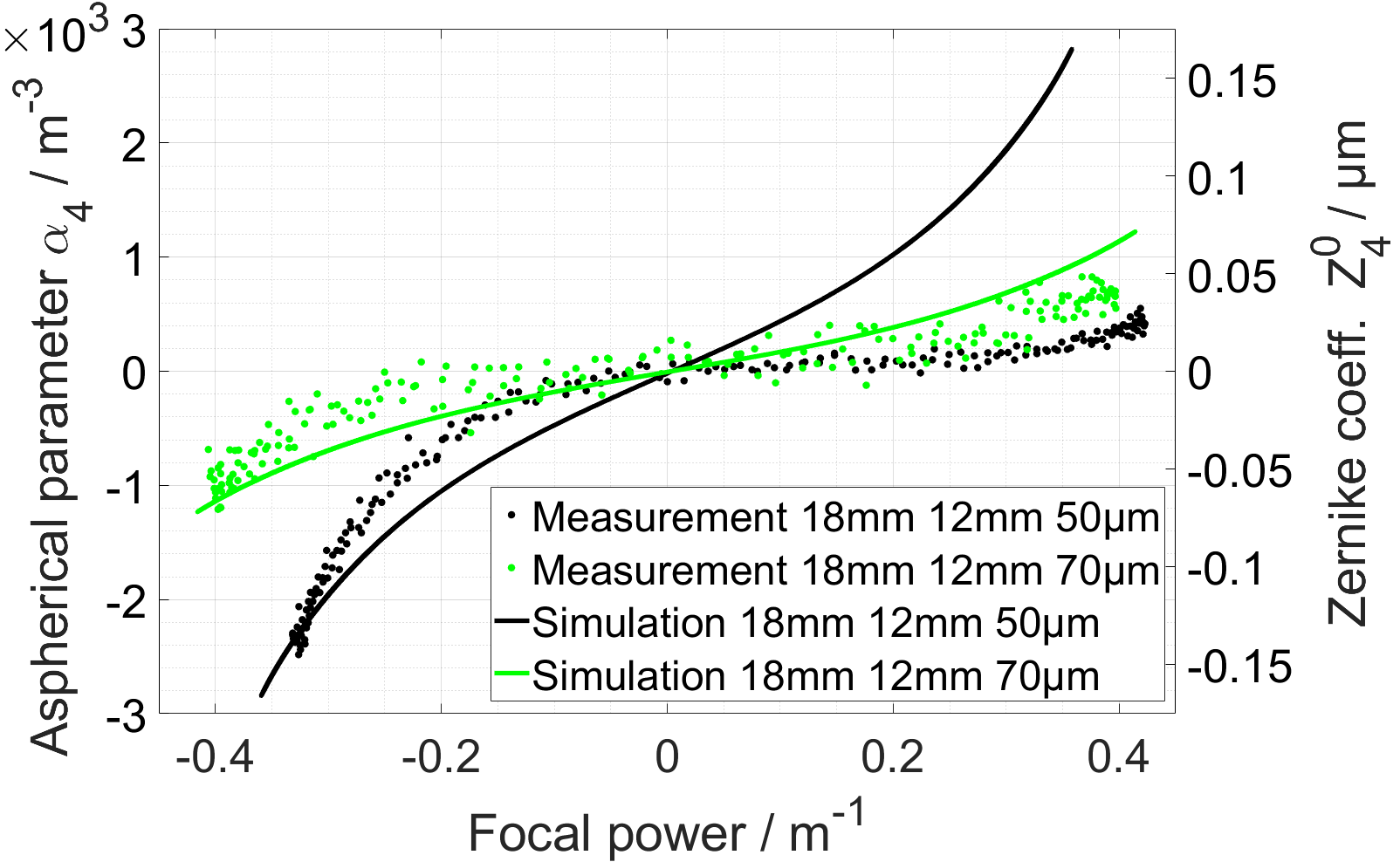}
	\includegraphics[width=0.49\linewidth]{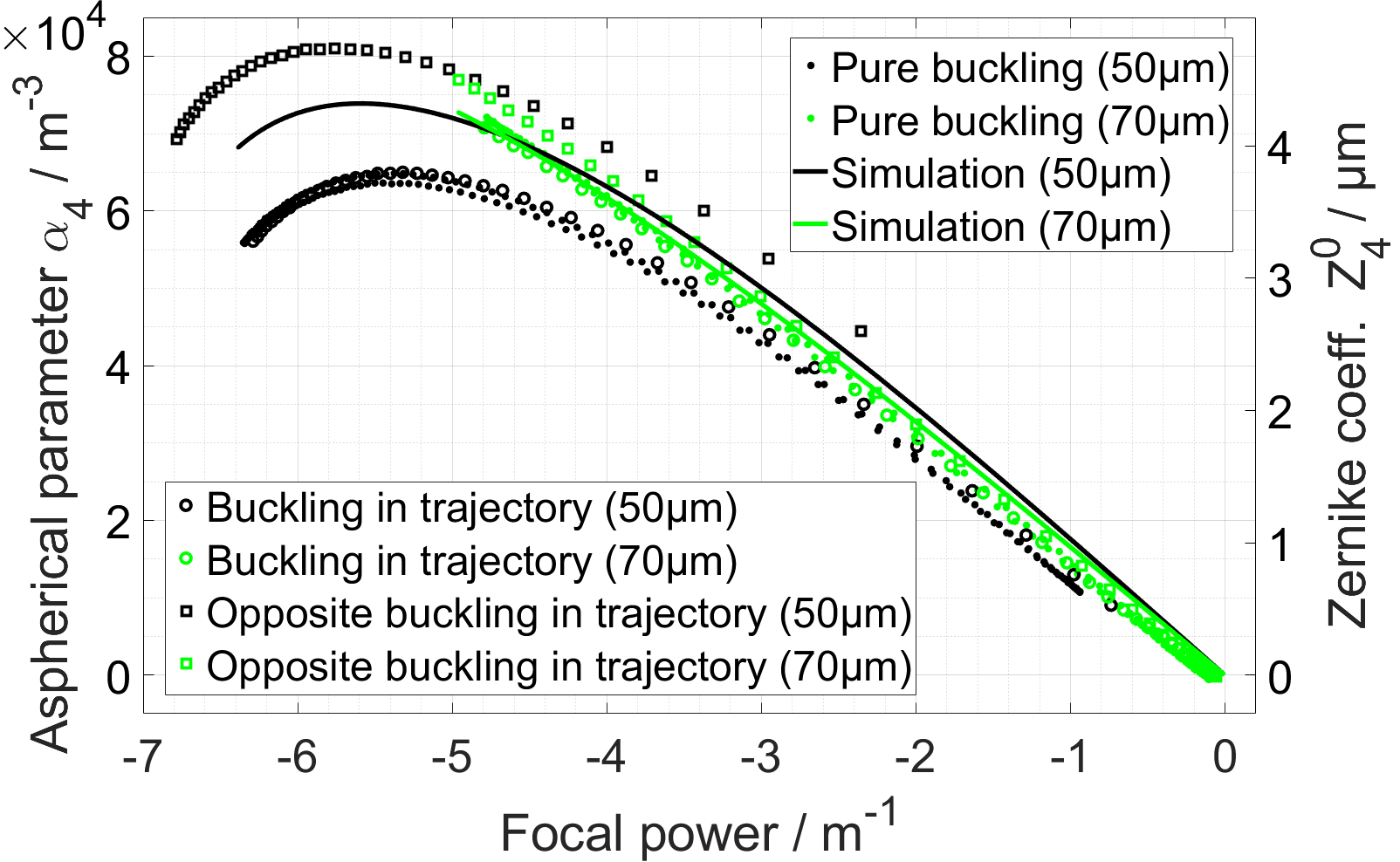}
	
	\caption{\myEmph{Simulated (\textit{solid}) and measured aspherical parameter $\alpha_4$ as a function of the focal power for the pure bending mode (\textit{dotted}, left), the pure buckling mode (\textit{dotted}, right), the buckling mode in the trajectory (\textit{circle}, right) and the buckling mode in the trajectory opposite to the pre-deflection (\textit{square}, right): Lenses with $D_{\text{out}}=\SI{18}{\milli\meter}$, $D_{\text{in}}=\SI{12}{\milli\meter}$ and $t_{\text{glass}}=\SI{50}{\micro\meter}$ (\textit{black}, \textit{blue}) and $t_{\text{glass}}=\SI{70}{\micro\meter}$ (\textit{green}). Opposite buckling measurement of trajectory mirrored at origin to compare influence of asymmetry (see \fref{fig:Sim_Traj} to find non-mirrored data).}}
	\label{fig:new}
\end{figure*}

\myEmph{In \fref{fig:new}, we show the aspherical parameter as a function of the focal power. For more clarity of presentation, we show only the \SI{50}{\micro\meter} membrane lens with the lower pre-deflection of the two measured \SI{50}{\micro\meter} membrane lenses. For the pure bending mode in the left graph, we find a good agreement with the simulation for the \SI{70}{\micro\meter} membrane (green), while there are lager deviations for positive focal powers in the \SI{50}{\micro\meter} membrane lens (black). The latter may be related to the larger pre-deflection of the \SI{50}{\micro\meter} membrane. For the thicker and stiffer \SI{70}{\micro\meter} membrane we do not see this behavior and find a reasonably good agreement with the simulation.}
\myEmph{The measurement of the pure buckling mode in \fref{fig:new} (\textit{dotted}, right graph) reproduces the shape of the simulation with a small shift for the \SI{70}{\micro\meter} membrane (green) and an even lager shift towards a smaller $\alpha_4$ for the \SI{50}{\micro\meter} membrane (black). This becomes also apparent in the buckling regions of the working range trajectory (see \fref{fig:Voltage_real_vs_sim}, right graph, 0.53-\SI{0.74}{\second} and 1.53-\SI{1.74}{\second}, where the same voltages are applied as for the pure buckling mode), where we find an asymmetry between the branches of positive and negative focal powers in particular for the \SI{50}{\micro\meter} lens. For both lenses, the simulation (\textit{solid line}), which does not have a pre-deflection, lies between the asymmetric measurements (\textit{circle, square}). The measurements of the ``opposite buckling direction" (inverse to the direction of pre-deflection) in the right graph of \fref{fig:new} were mirrored at the origin for better comparison.}

\begin{figure*}[tbph!]
	\centering
	\includegraphics[width=0.98\linewidth]{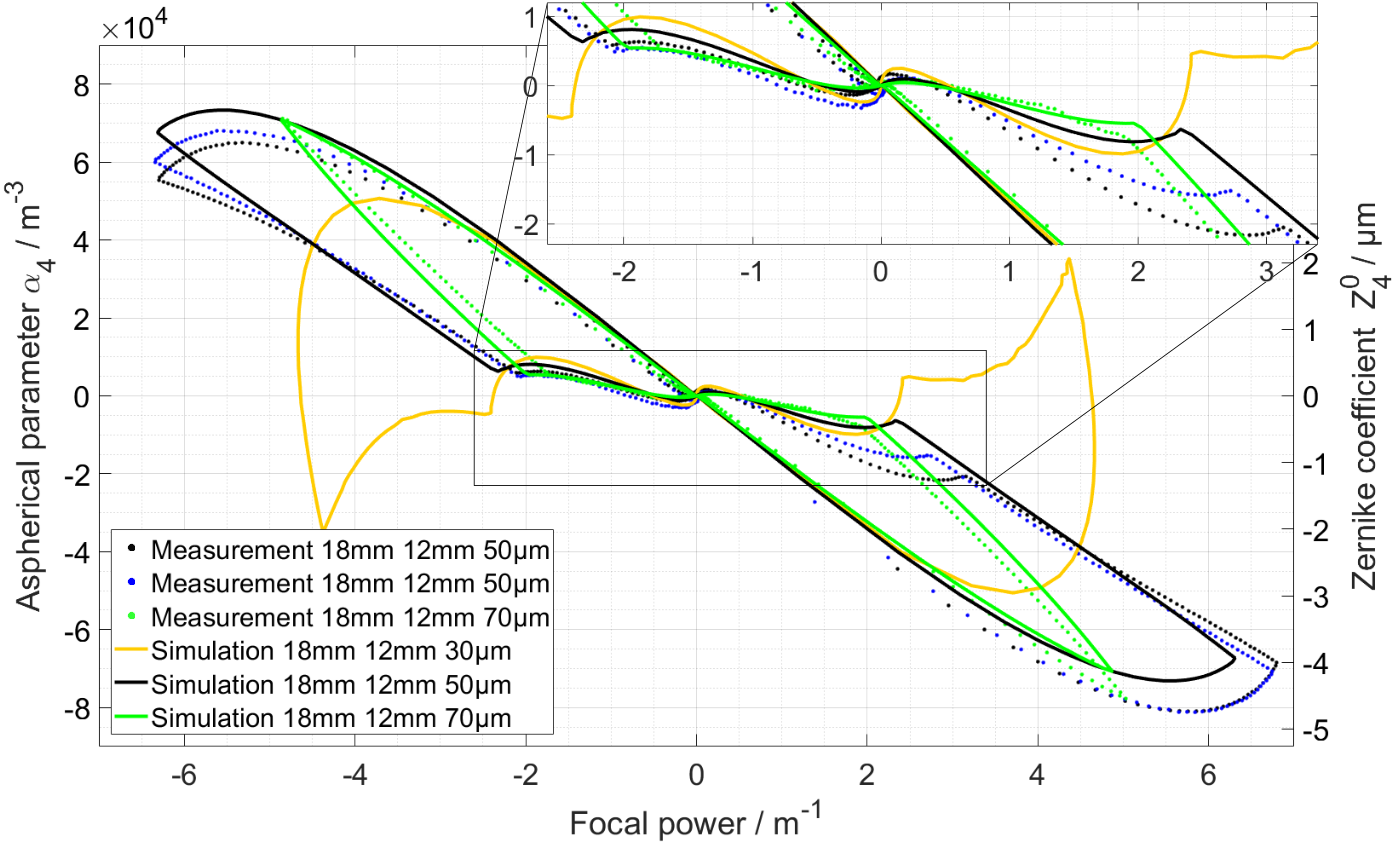}
	\caption{Comparison of the simulated (\textit{solid}) and measured (\textit{dotted}) working range of the lenses with $D_{\text{out}}=\SI{18}{\milli\meter}$, $D_{\text{in}}=\SI{12}{\milli\meter}$ and $t_{\text{glass}}=\SI{50}{\micro\meter}$ (\textit{black}, \textit{blue}) and $t_{\text{glass}}=\SI{70}{\micro\meter}$ (\textit{green}) with detailed view for low focal powers as an inset. Simulation of a lens with $t_{\text{glass}}=\SI{30}{\micro\meter}$ shown in \textit{yellow}.}
	\label{fig:Sim_Traj}
\end{figure*}

In \fref{fig:Sim_Traj}, we show the working range of the focal power and aspherical coefficient for the voltage trajectory described in section \ref{sec:sim} and the corresponding simulation result. Again, the simulation reproduces the behavior of these two different designs reasonably well with a small asymmetry towards positive focal powers and smaller aspherical parameters for both membranes. 
Also here, we see that the two lenses with the \SI{50}{\micro\meter} design (black, (a) and blue, (b)) agree in their focal power and aspherical ranges and show only small deviations, e.g. due to pre-deflection for zero electric field (see \fref{fig:Sim_Bend_Buck} left).

We also fabricated a lens with a \SI{30}{\micro\meter} thick membrane, but it showed a very strong pre-deflection of \SI{0.35}{\per\meter} and also a strong asymmetry as the thin membrane is difficult to handle and is more affected by uneven piezo sheets and fabrication imperfections, so we did not include it in the data analysis. However, we see in \fref{fig:Sim_Traj} that the simulation results for the \SI{30}{\micro\meter} membrane lens promise a great improvement in the aspherical tuning region.

\begin{figure*}[bhpt!]
	\centering
	\includegraphics[width=0.49\linewidth]{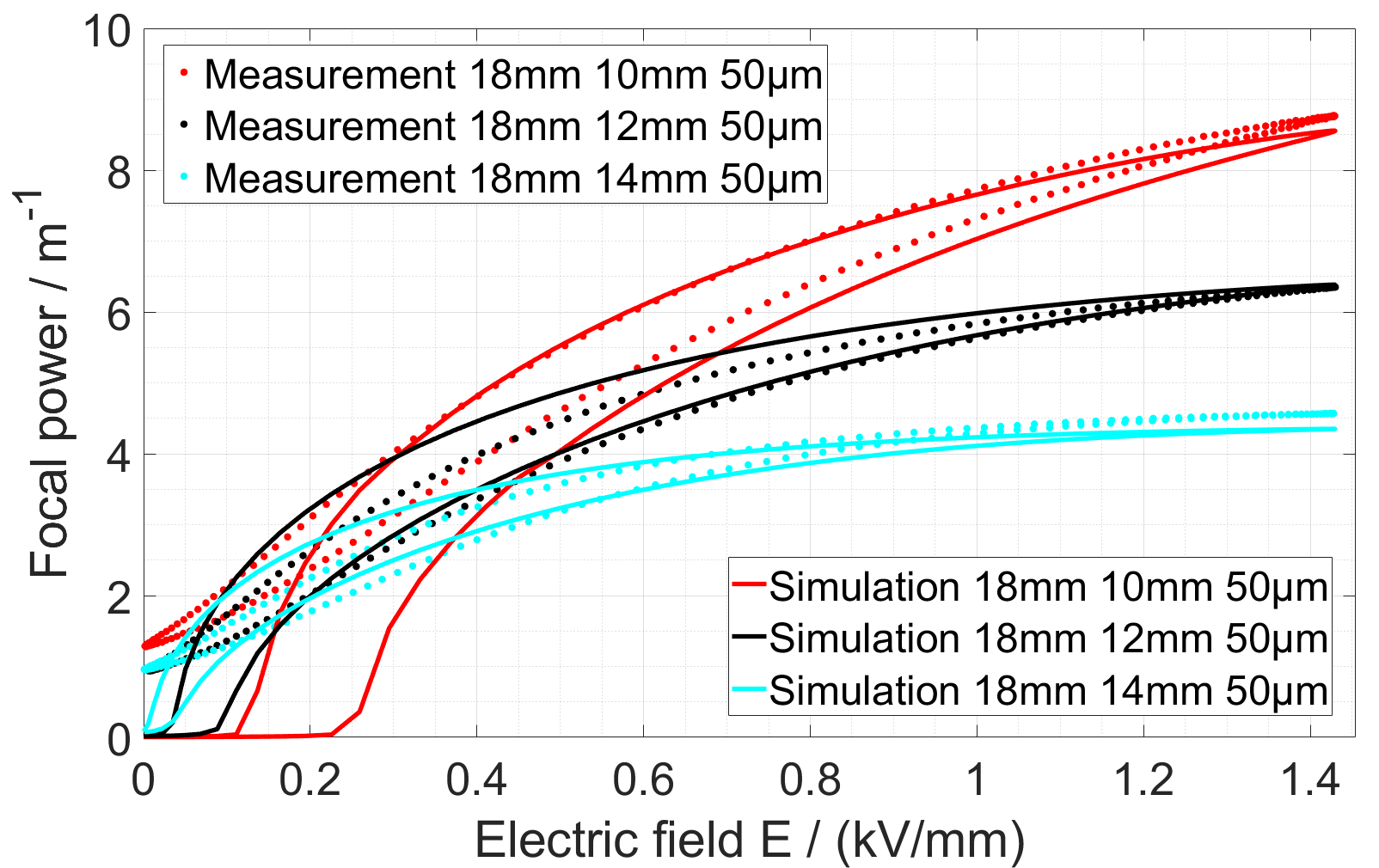}
	\includegraphics[width=0.49\linewidth]{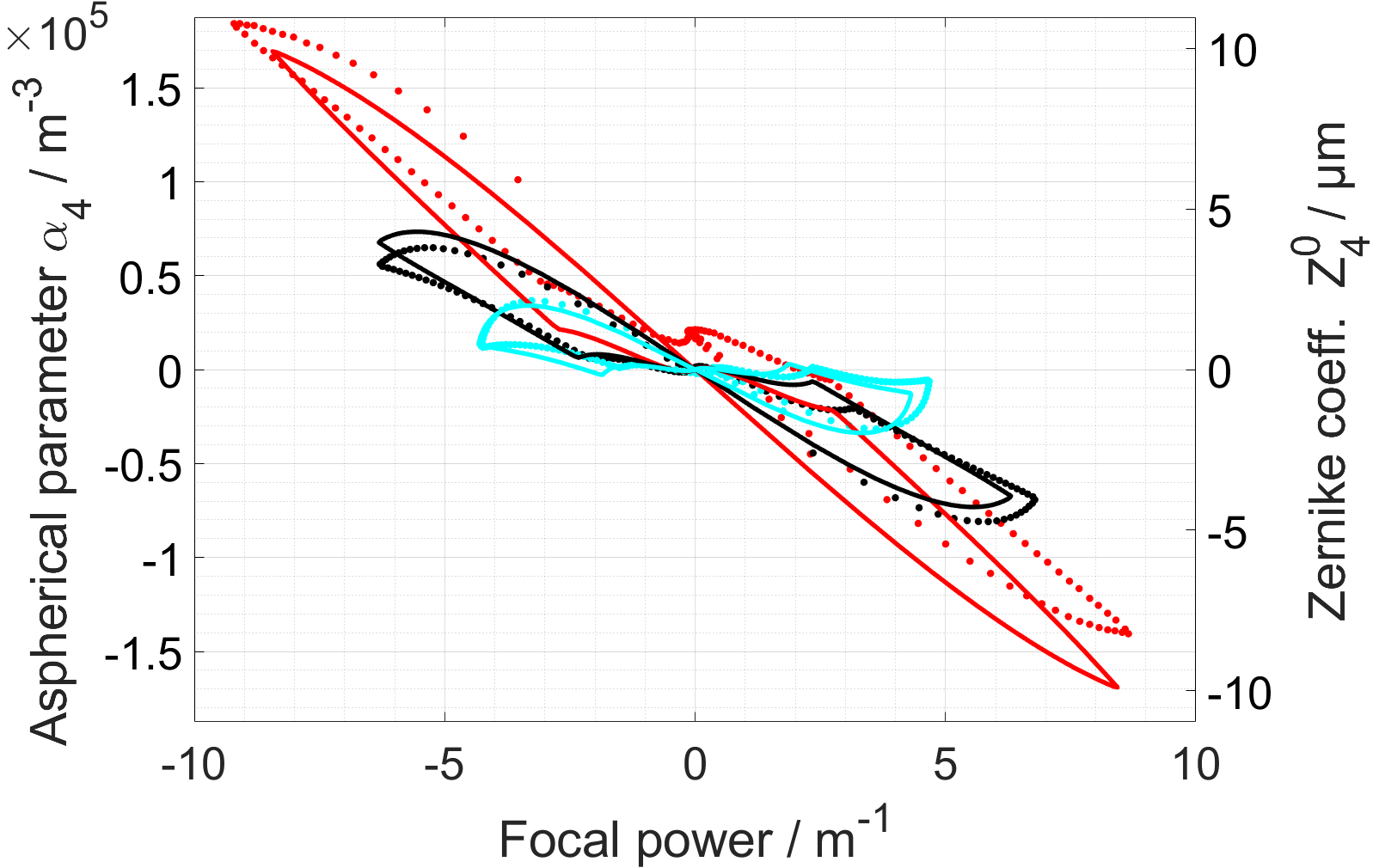}
	\caption{Comparison of measurement (\textit{dotted}) and simulation (\textit{solid}) for a varying inner diameter $D_{\text{in}}$ of the piezo rings (\SI{10}{}, \SI{12}{} and \SI{14}{\milli\meter}).  Focal power in the buckling mode (left) and aspherical parameter and focal power in the maximal trajectory (right).
	}
	\label{fig:Inner}
\end{figure*}

\begin{figure*}[bhpt!]
	\centering
	\includegraphics[width=0.49\linewidth]{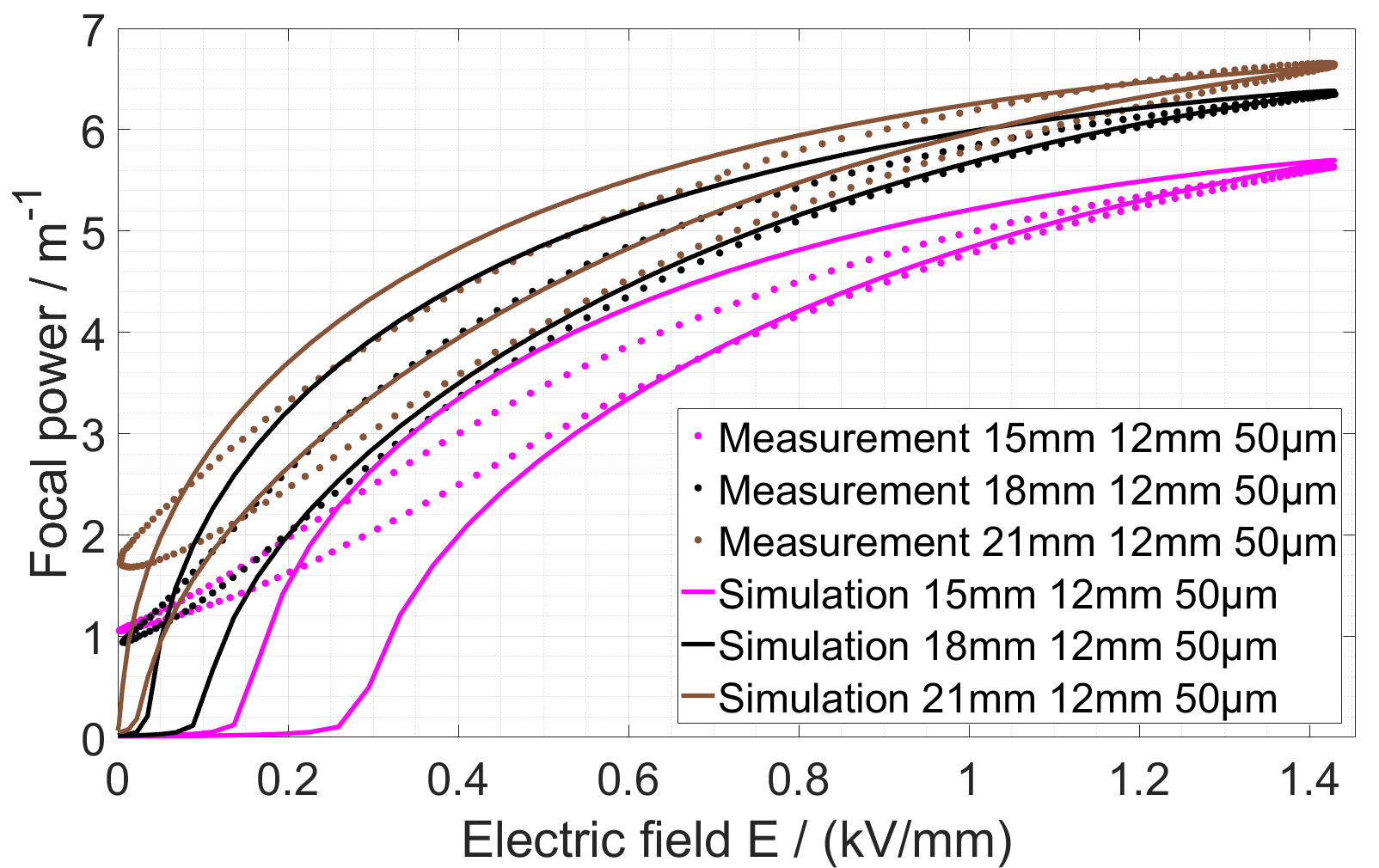}
	\includegraphics[width=0.49\linewidth]{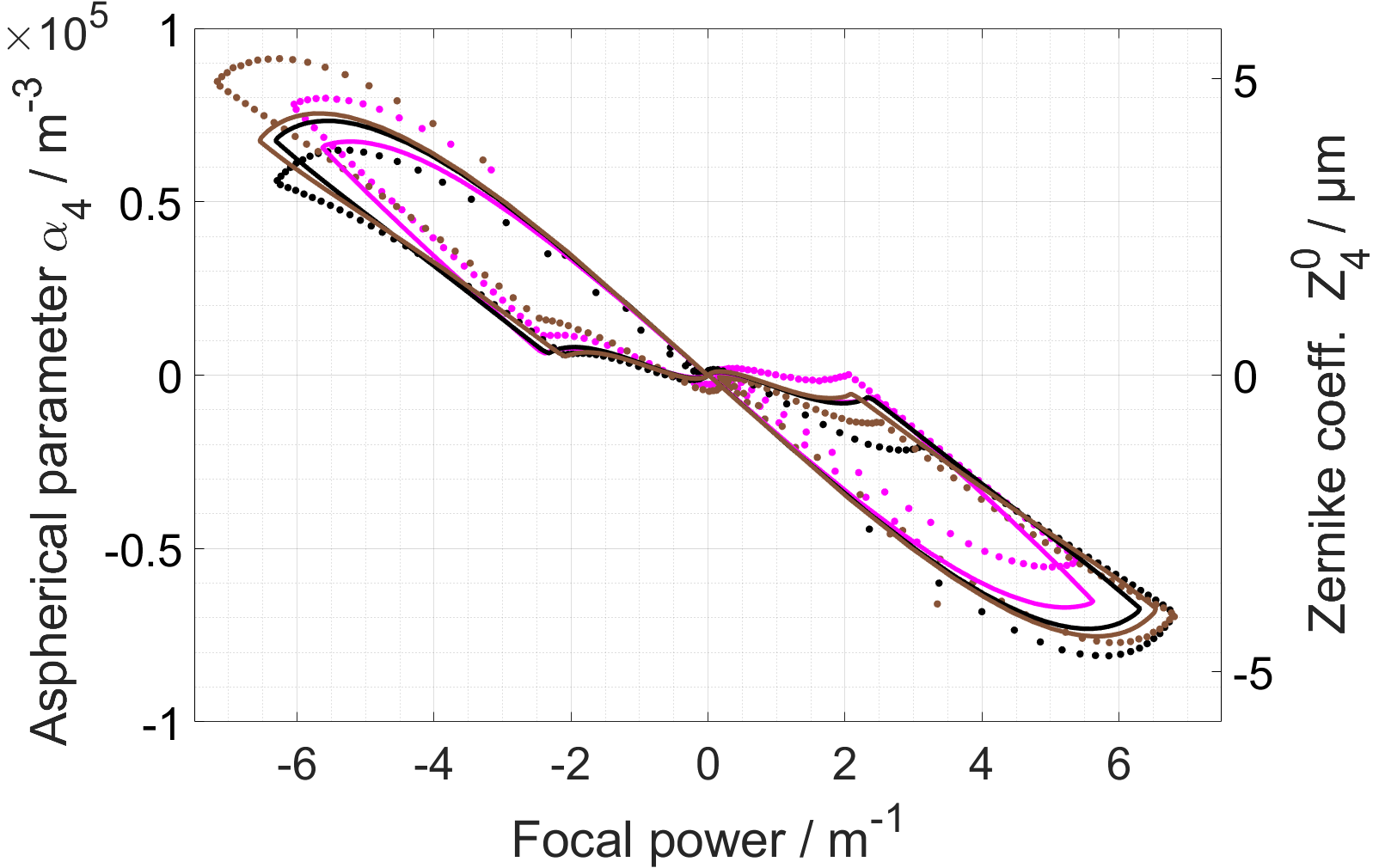}
	\caption{Comparison of measurement (\textit{dotted}) and simulation (\textit{solid}) for a varying outer diameter $D_{\text{out}}$ of the piezo rings (\SI{15}{}, \SI{18}{} and \SI{21}{\milli\meter}).}
	\label{fig:Outer}
\end{figure*}

Reducing the inner diameter, however, increases the maximal focal power range as shown in \fref{fig:Inner}, in agreement with \eref{eq:buck} (see \tref{tab:coeffs}). The deviation between the analytical estimate and the experimental result decreases with increasing inner diameter (and fixed outer diameter) as the amount of piezo material increases and hence the effect of the glass membrane stiffness decreases. There is also a small increase in the offset electric field for decreasing inner diameters, as smaller inner diameters result in a smaller aspect ratio of the glass membrane and hence more resistance to buckling. We define the aspect ratio as the width of the glass membrane $D_{\text{in}}$ divided by its thickness $t_{\text{glass}}$. The simulation matches very well for all the designs taking into account the asymmetry of the measurements. In the trajectory on the right graph, we see that, while the focal power increases with smaller $D_{\text{in}}$, the aspherical tuning range shifts to more hyperbolic values (lager $\alpha_4$ values). This comes from the fact that the wider piezo rings resist more the bending deformation of the glass membrane, so the lens profile becomes more hyperbolic. While we see a relatively strong deviation for the smallest aperture, this one also had some fabrication asymmetry (vertical offset).
Finally, when changing the outer diameter (\fref{fig:Outer}), we find that a reduction of the outer diameter, i.e., a smaller amount of active piezo material causes less available force and therefore less buckling deflection (magenta). Secondly, while the trend is not as clear as for the different inner diameters, the wider rings seem to result also in larger aspherical parameters, i.e., a more hyperbolic behavior (brown).

In \tref{tab:coeffs}, we show an overview of the fitted parameters. We see that the linear fit of the bending mode and the analytical estimate deviate by a factor of approximately~2\myEmph{, beacuse we assume in all estimates a typical large-field charge coefficient $d_{31}=\SI{-487}{\pico\meter\over\volt}$ \cite{BinalSMS}, which does not apply for small and negative electric fields and results in overestimated analytical values in the bending mode as already explained in section~\ref{sec:fab}. Additionally, for the bending mode, we assumed in section~\ref{sec:op} neutral mechanical planes in the centers of the piezo rings, which are probably shifted due to the mechanical behavior of the piezo-glass composite. Also the assumption of spherical displacements, the neglection of forces and the assumed boundary conditions, as explained in section~\ref{sec:op}, may explain the larger values in the analytical approximation of the bending and buckling mode.} The different geometries show relatively little variations, with the exception of lens 7 which in fact had an atypical asymmetric behavior in the bending mode with a large pre-deflection. The relatively (compared to the estimate) large coefficient of the \SI{70}{\micro\meter} membrane may result from a better transfer of the deformation from the boundary to the center in the thicker membrane and the smaller curvature of the \SI{14}{\milli\meter} lens may have the same reason, that the lens is bent at the side but remains relatively flat at the center.

We have a similar approximate factor in the buckling coefficient. Here, we see the clear correlation with the inner diameter as predicted by \eref{eq:buck} and a small trend towards larger coefficients, the wider the piezo ring becomes ($(D_{\text{out}}-D_{\text{in}}) / 2$, compare lens 6 and~7). 

The offset voltage shows first of all negative values, which are simply due to the strain in the piezo caused by the long-term creep (\fref{fig:Voltage_real_vs_sim}). The various values can then be explained with two effects: A smaller aspect ratio of the membrane (lenses 3 and 4) results in an increased buckling threshold as expected from plate theory. Secondly, an increasing amount of piezo material (compare lens 1 to lenses 6 and 7) decreases the offset because it creates larger forces to overcome the critical buckling load.
\begin{table}[h!]
\centering
\small
\caption{\myEmphThree{Overview over the fitted coefficients of the measured lenses}}
\label{tab:coeffs}
\begin{tabular}{|l|c|c|c|c|c|c|}\hline
Lens  &$D_{\text{out}}$/$D_{\text{in}}$/$t_{\text{glass}}$ & \multicolumn{2}{c|}{Bending coefficient $a$} & \multicolumn{2}{c|}{Buckling coefficient $b$} &  $E_{0}$\\
\cline{3-7}
&in& analytical &meas. & analytical&meas. &meas.\\
&$\SI{}{\milli\meter}$/$\SI{}{\milli\meter}$/$\SI{}{\micro\meter}$&/$\SI[per-mode = fraction]{}{\per\volt}$&/$\SI[per-mode = fraction]{}{\per\volt}$&/ $\sqrt{\SI[per-mode = fraction]{}{\per\meter\per\volt}}$ &/$\sqrt{\SI[per-mode = fraction]{}{\per\meter\per\volt}}$&/\SI[per-mode = fraction]{}{\kilo\volt\per\milli\meter}\\
\hline
\hline
\textbf{1} (a)     & 18                / 12              / 50  			  & \SI{2.28e-6}{}&  \SI{1.02e-6}{} &\color[rgb]{0,0,0} \SI{9.01e-3}{} &\color[rgb]{0,0,0}\SI{5.60e-3}{} & \SI{-0.07}{} \\
\color[rgb]{0,0,1}\textbf{1} (b)     & 18                / 12              / 50  			  & \SI{2.28e-6}{}&  \SI{1.09e-6}{} &\color[rgb]{0,0,0} \SI{9.01e-3}{} &\color[rgb]{0,0,0}\SI{5.69e-3}{} & \SI{-0.07}{} \\
\hline
\color[rgb]{1,0.8,0}\bf2 \,     & \textit{18}                / \textit{12}              / \textit{\textbf{30}} & \SI{2.53e-6}&\textit{-}		& \SI{9.01e-3}{}&\textit{-} & \textit{-}\\
\color[rgb]{0,0.58,0}\bf3 \,    & 18                / 12              / \textbf{70} & \SI{2.08e-6}{}&  \SI{1.06e-6}{}& \SI{9.01e-3}{}&\SI{5.77e-3}{} & \SI{0.75}{} \\
\hline
\color[rgb]{1,0,0}\bf4  \,   & 18                / \textbf{10}       / 50  			& \SI{2.28e-6}{}&  \SI{0.99e-6}{} &\color[rgb]{0,0,0} \SI{10.81e-3}{}&\color[rgb]{0,0,0}\SI{7.92e-3}{} & \SI{0.09}{}\\
\color[rgb]{0,1,1}\bf5 \,    & 18                / \textbf{14}       / 50  			& \SI{2.28e-6}{}&  \SI{0.89e-6}{} &\color[rgb]{0,0,0} \SI{7.72e-3}{}&\color[rgb]{0,0,0}\SI{3.85e-3}{} & \SI{-0.30}{}\\
\hline
\color[rgb]{1,0,1}\bf6 \,    & \textbf{15}         / 12              / 50  			& \SI{2.28e-6}{}&  \SI{1.02e-6}{} & \SI{9.01e-3}{}& \SI{5.22e-3}{} & \SI{0.12}{}\\
\color[rgb]{0.53,0.33,0.22}\bf7	\,		& \textbf{21}         / 12              / 50  			& \SI{2.28e-6}{}&  \SI{0.69e-6}{} & \SI{9.01e-3}{}& \SI{5.60e-3}{} & \SI{-0.12}{}\\
\hline
\end{tabular}
\end{table} 
\subsection{Effects of the fabrication}
In the above simulations, we included some estimated mean fabrication parameters: The thickness of the glue layer ($\SI{25}{\micro\meter}$), the glue flowing onto the membrane ($\SI{130}{\micro\meter}$) and the thermal stress of the gluing process (with $\alpha_{piezo}=\SI{6e-6}{\per\kelvin}$). As they cannot be controlled perfectly during fabrication, it is important to know how they affect the behavior of the lens.

\begin{figure*}[bhpt!]
	\centering
	\includegraphics[width=0.32\linewidth]{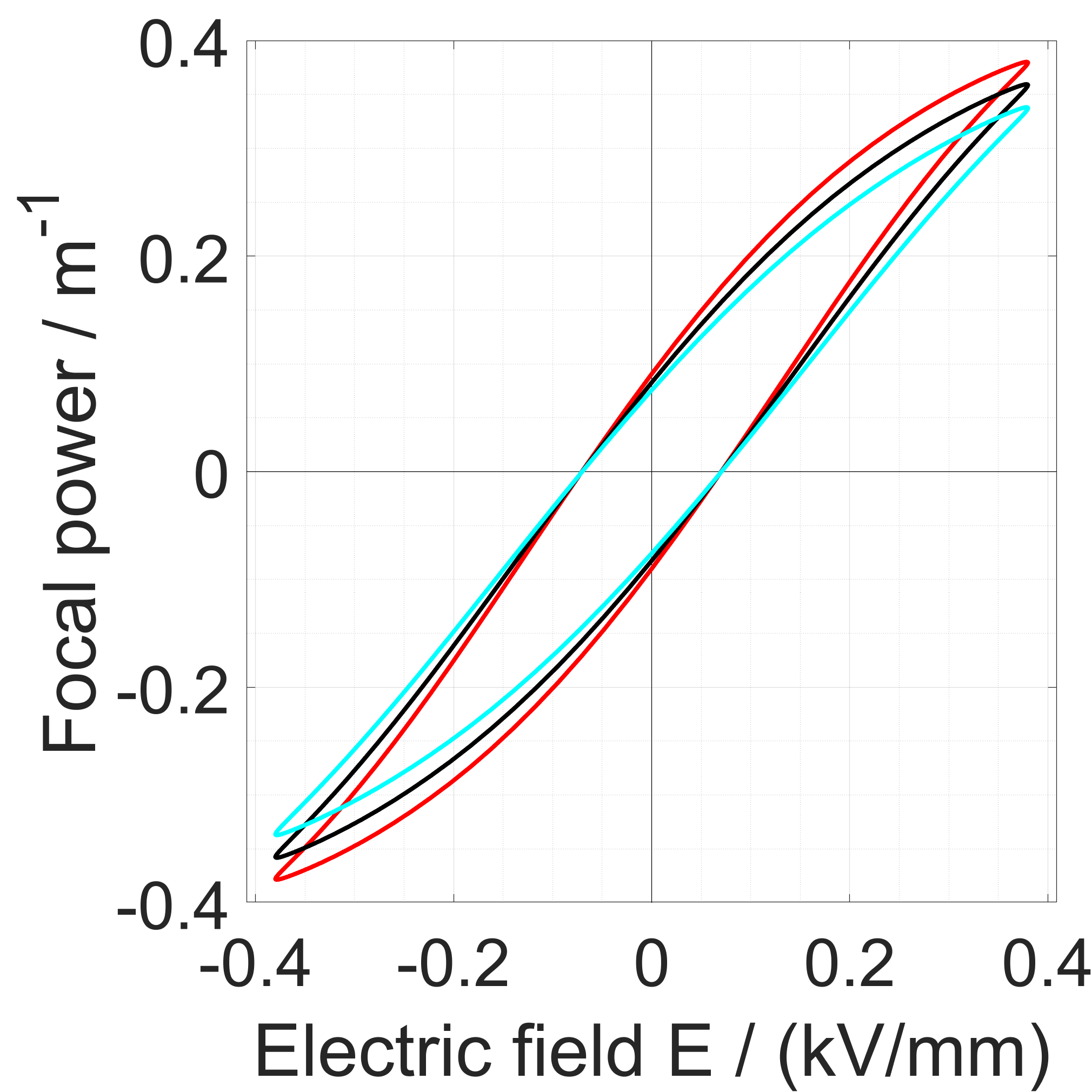}
	\includegraphics[width=0.32\linewidth]{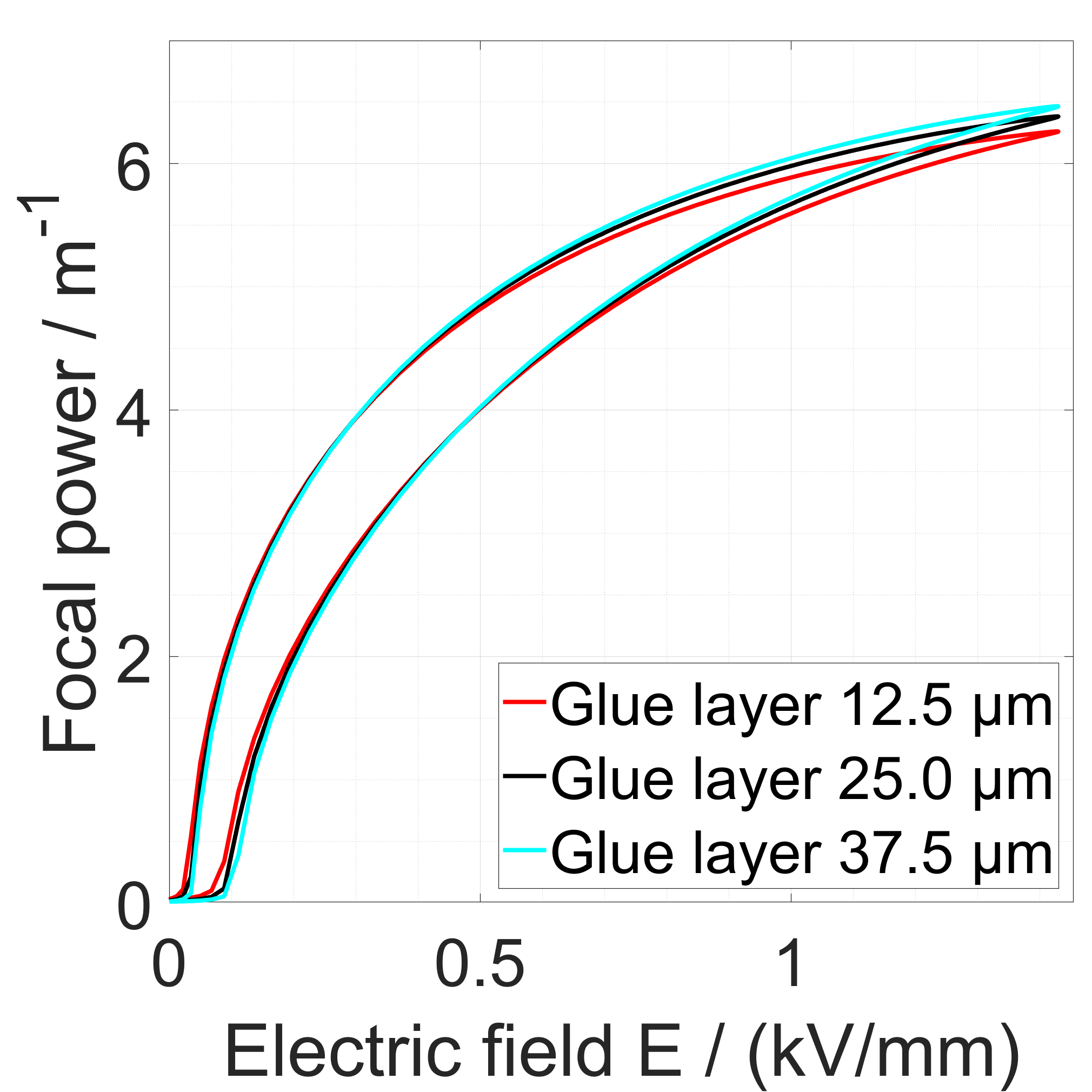}
	\includegraphics[width=0.32\linewidth]{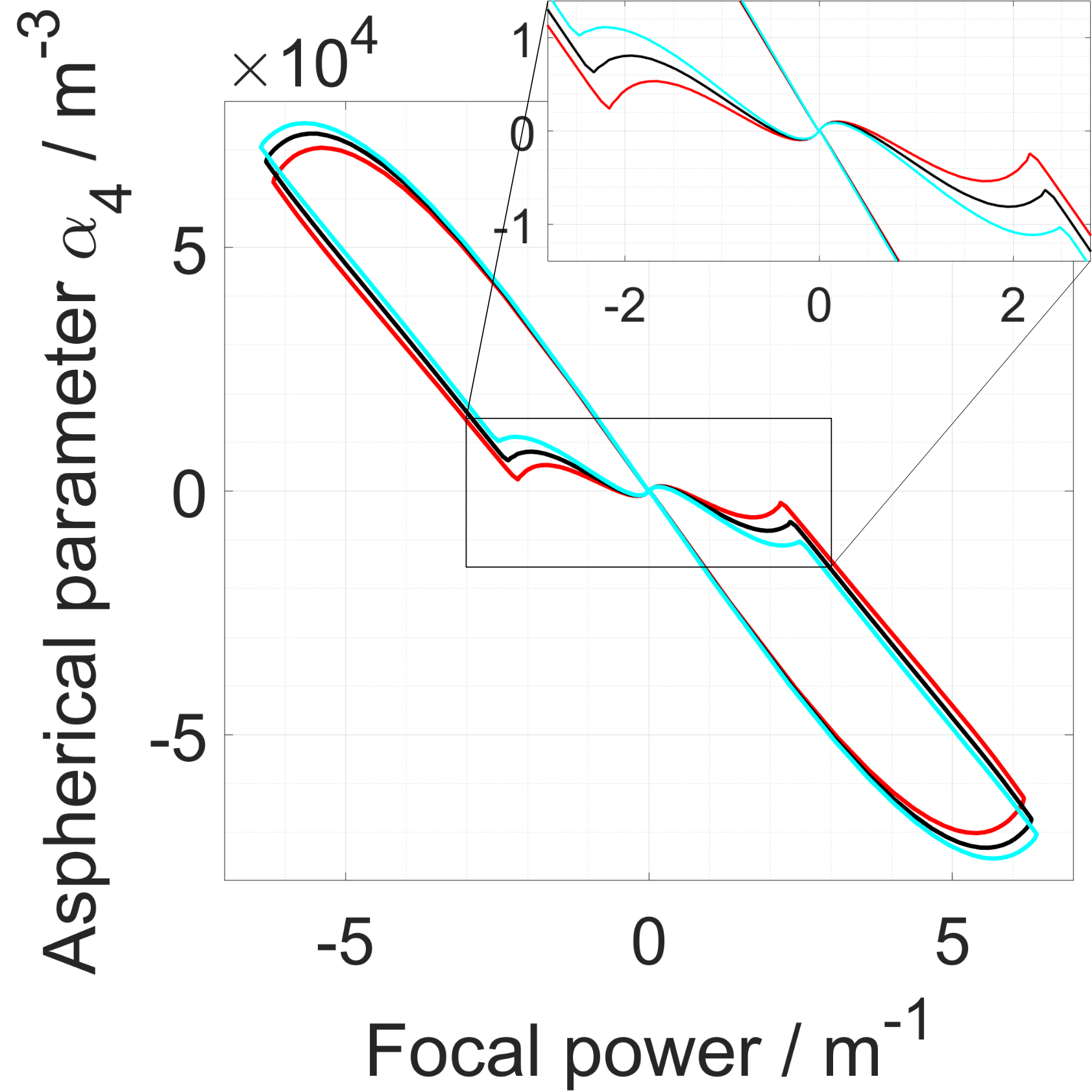}
	\caption{Effect of glue layer thickness variation for the bending mode (left), buckling mode (center) and aspherical working range (right).}
	\label{fig:gluelayer}
\end{figure*}

In \fref{fig:gluelayer} we find that a thinner glue layer results in a higher focal power in the bending mode as expected from \eref{eq:bend} where a thinner glue layer reduces the distance $s$ of the piezos. It also increases the aspherical tuning range for small focal powers. On the other hand, a thinner glue layer decreases the focal power in the buckling mode and also decreases the working range for the focal power and the aspherical tuning (see \fref{fig:gluelayer} right).

\begin{figure*}[bhpt!]
	\centering
	\includegraphics[width=0.32\linewidth]{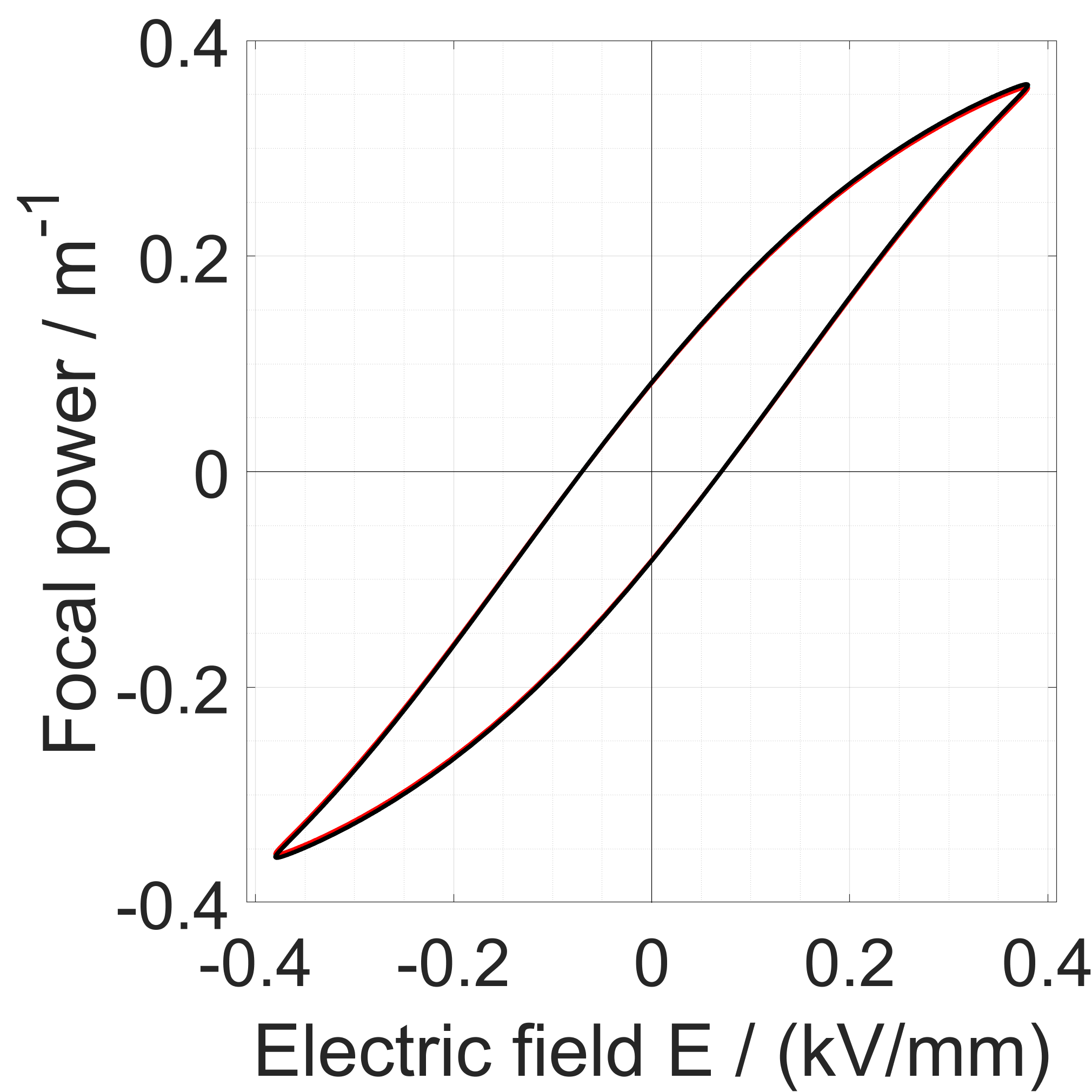}
	\includegraphics[width=0.32\linewidth]{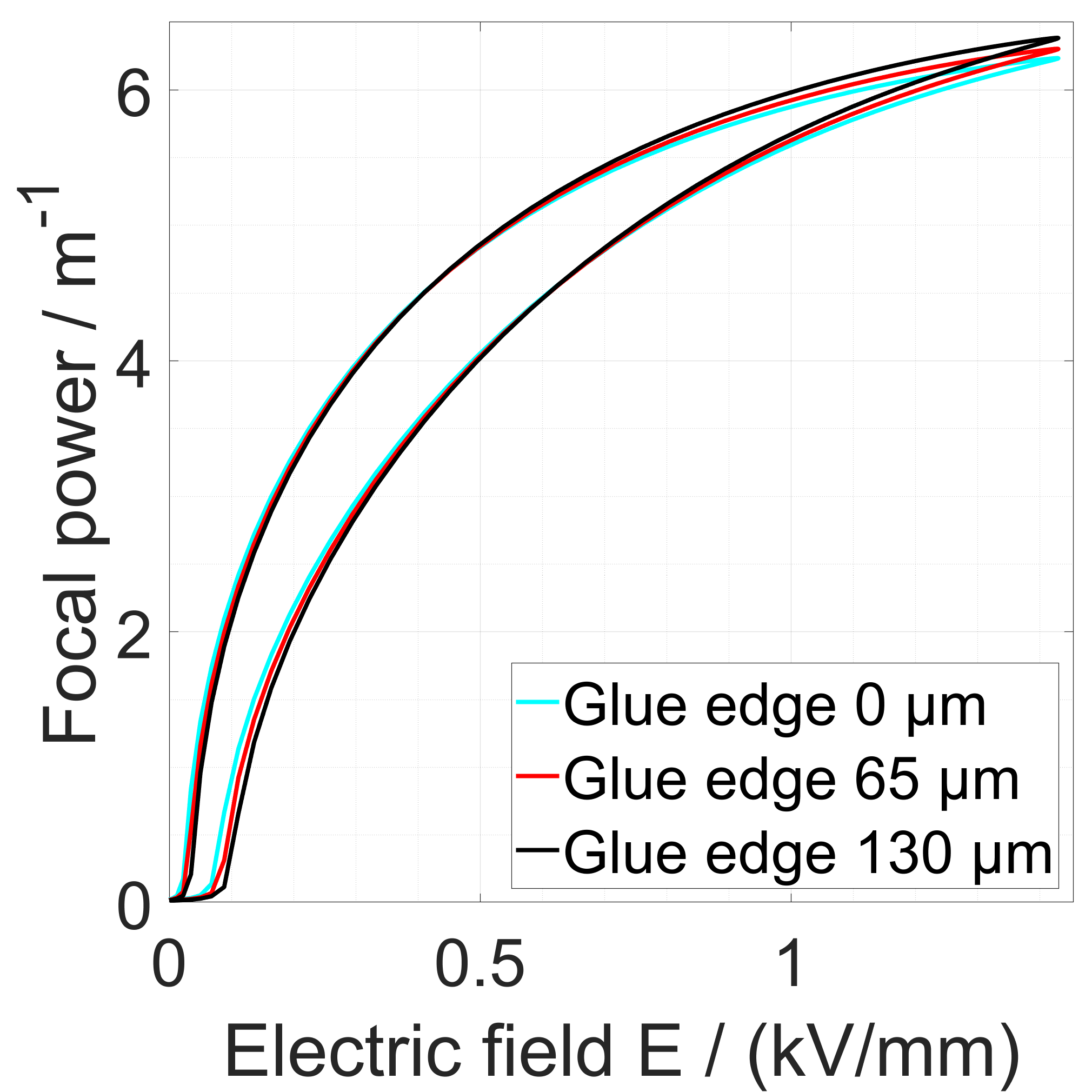}
	\includegraphics[width=0.32\linewidth]{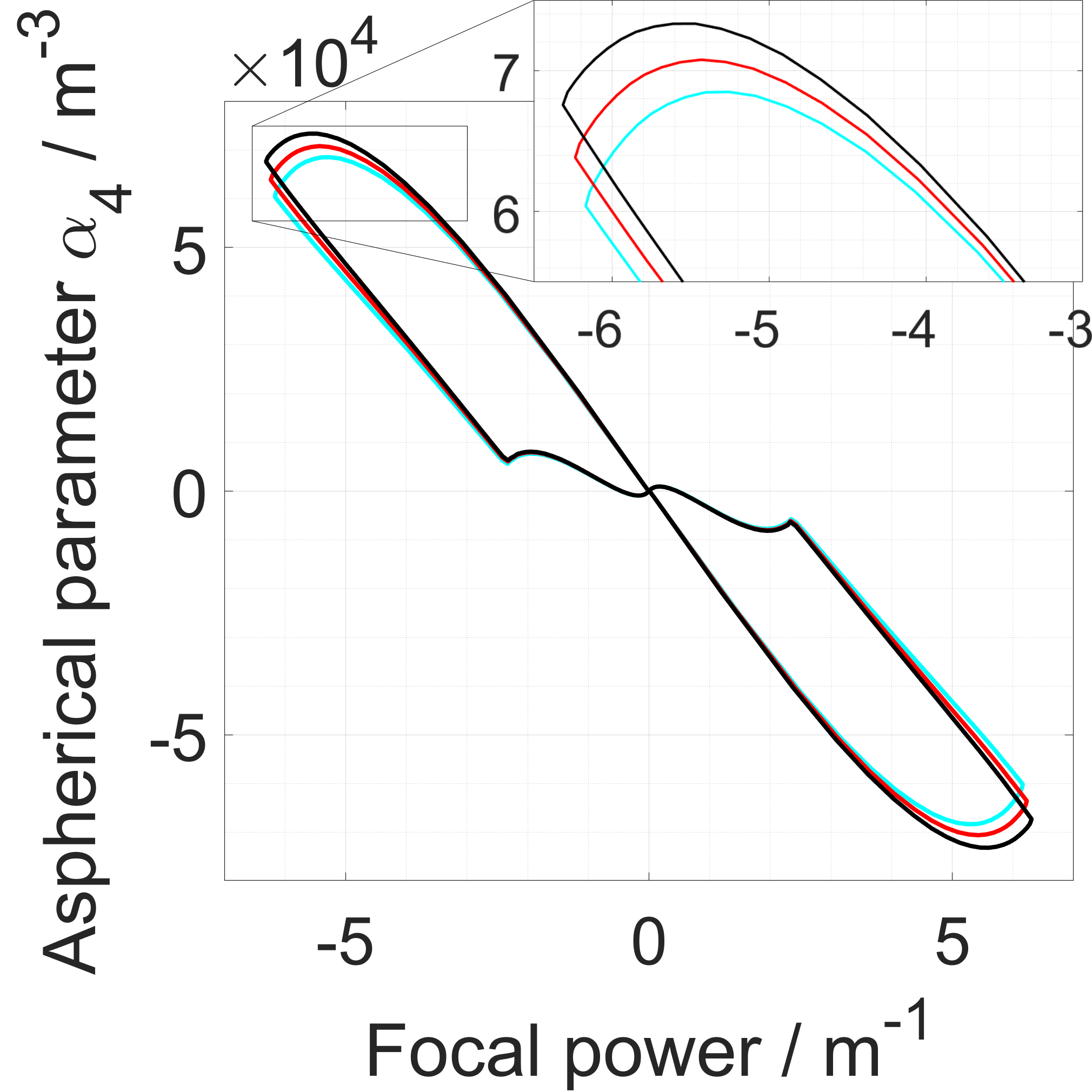}
	\caption{Effect of the glue edge (value of width and height) for the bending mode (left), buckling mode (center) and aspherical working range (right)}
	\label{fig:glueedge}
\end{figure*}

In \fref{fig:glueedge} we find that an increase of the glue edge, caused by additional glue flowing out of the glue layer while bonding, has nearly no effect on the bending mode. However, it increases the focal power of the buckling mode as a wider glue layer has a similar effect as a decrease of the inner piezo diameter $D_{\text{piezo}}$, resulting in a higher focal power by \eref{eq:buck}. There is similarly a small increase in the aspherical tuning range with increasing glue edge as shown in the right graph of \fref{fig:glueedge}. When taking into consideration that the lens is a full 3-dimensional setup, an asymmetry of the glue edge also results in an asymmetry of the surface curvature and therefore higher asymmetrical aberrations. Hence, a glue edge should be avoided or at least reduced to a minimum.

\begin{figure*}[bhpt!]
	\centering
	\includegraphics[width=0.32\linewidth]{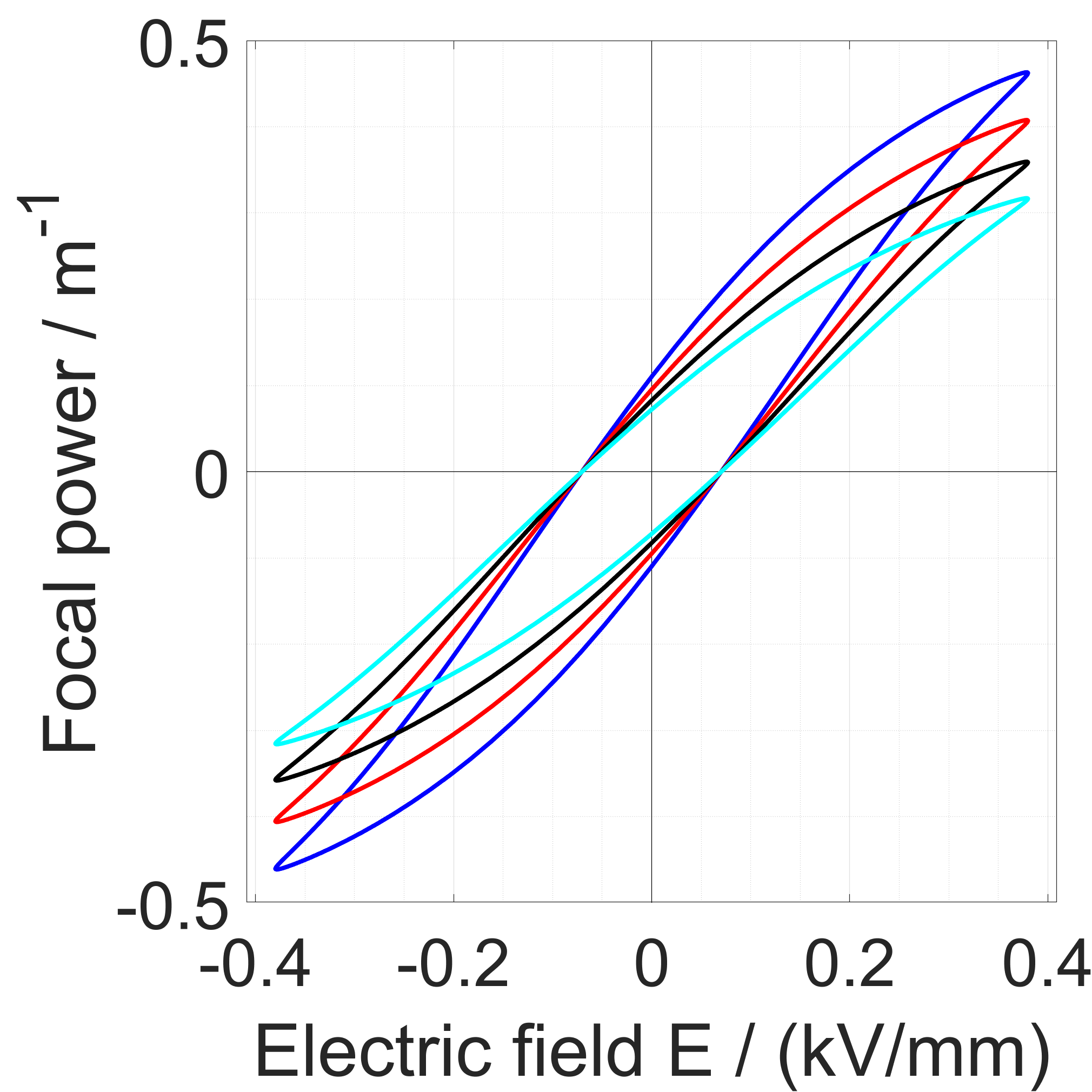}
	\includegraphics[width=0.32\linewidth]{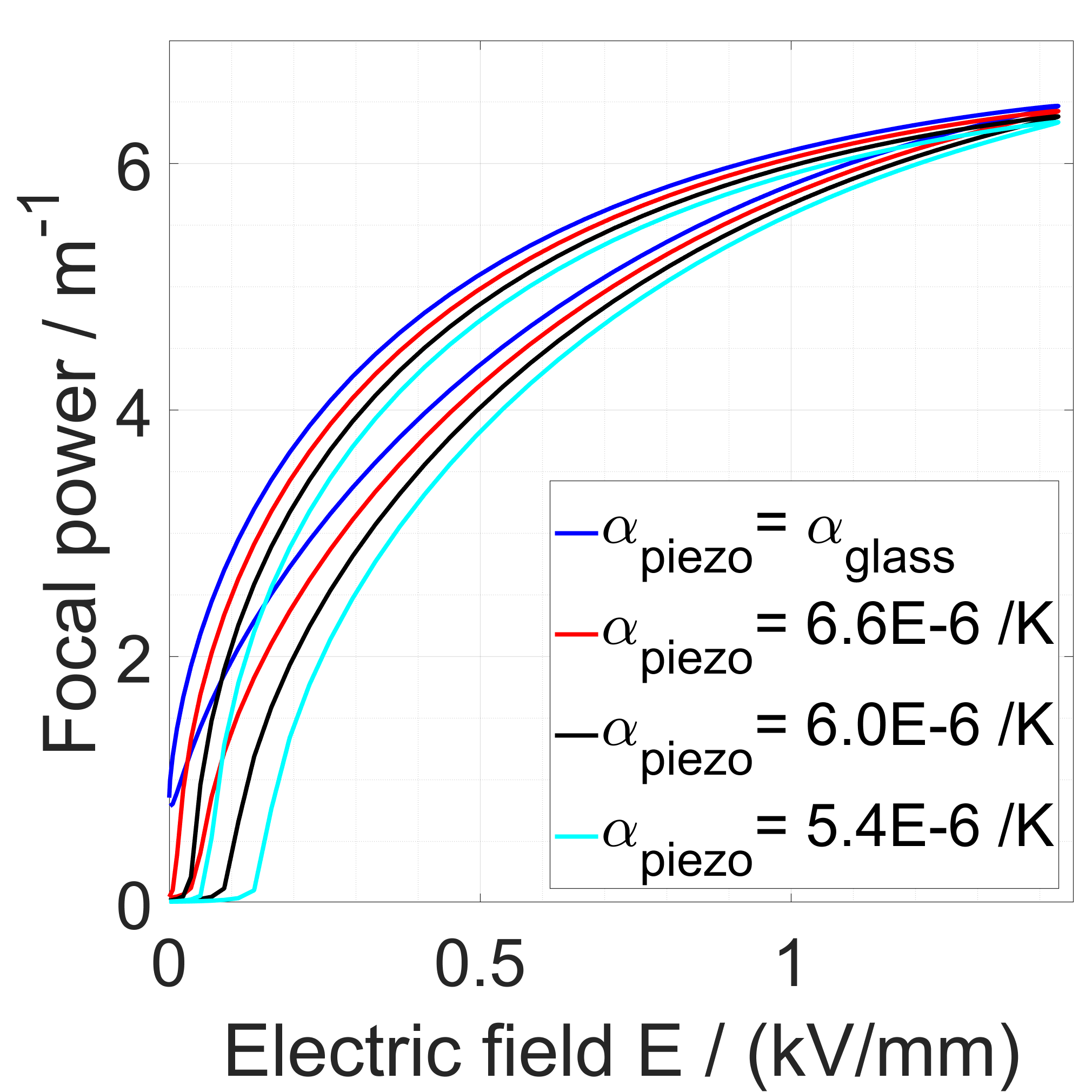}
	\includegraphics[width=0.32\linewidth]{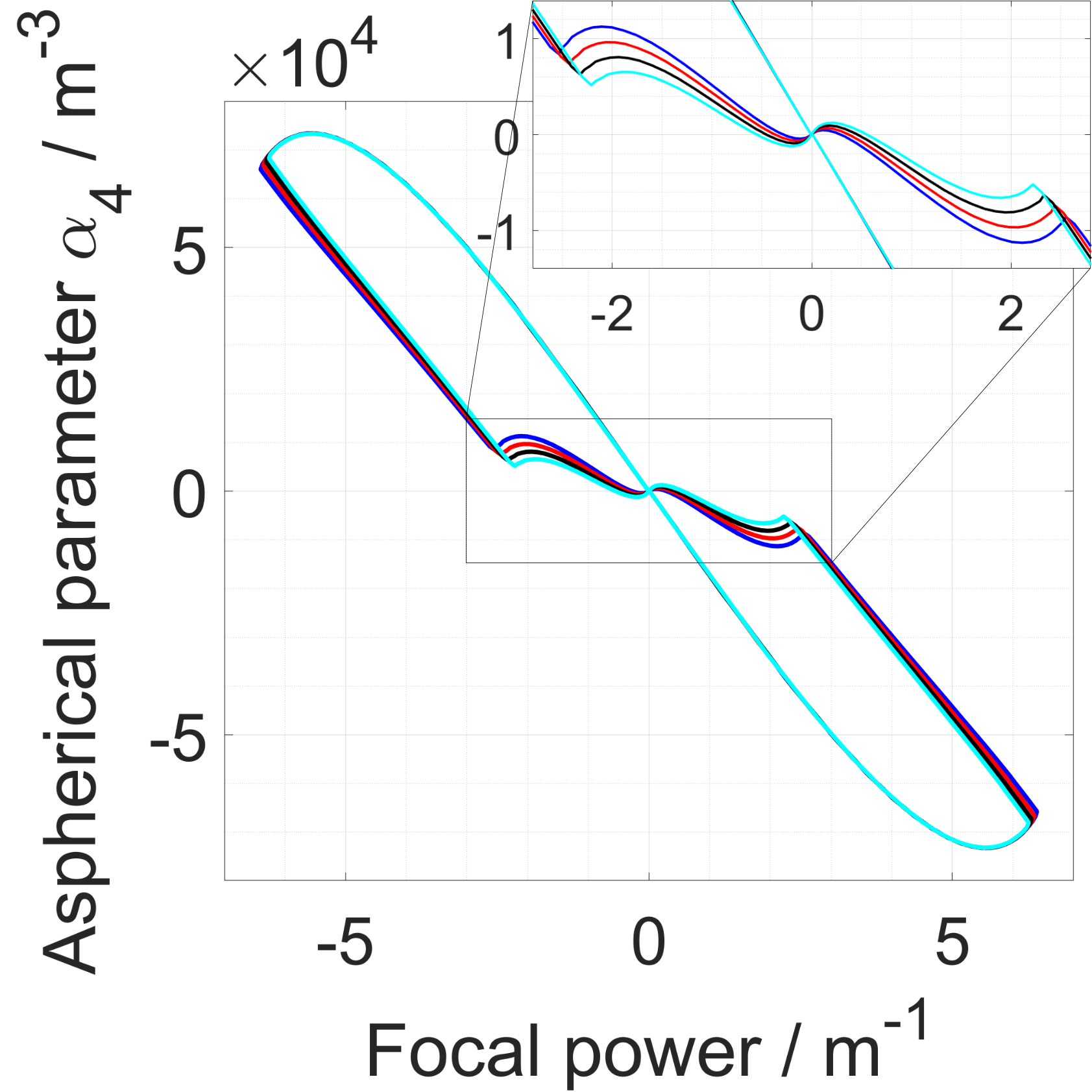}
	\caption{Effect of the thermal stress caused by different thermal coefficients and a curing temperature \SI{26}{\degree} above the measurement temperature for the bending mode (left), buckling mode (center) and aspherical working range (right).}
	\label{fig:thermalstrain}
\end{figure*}

\begin{figure*}[bhpt!]
	\centering
	\includegraphics[width=0.49\linewidth]{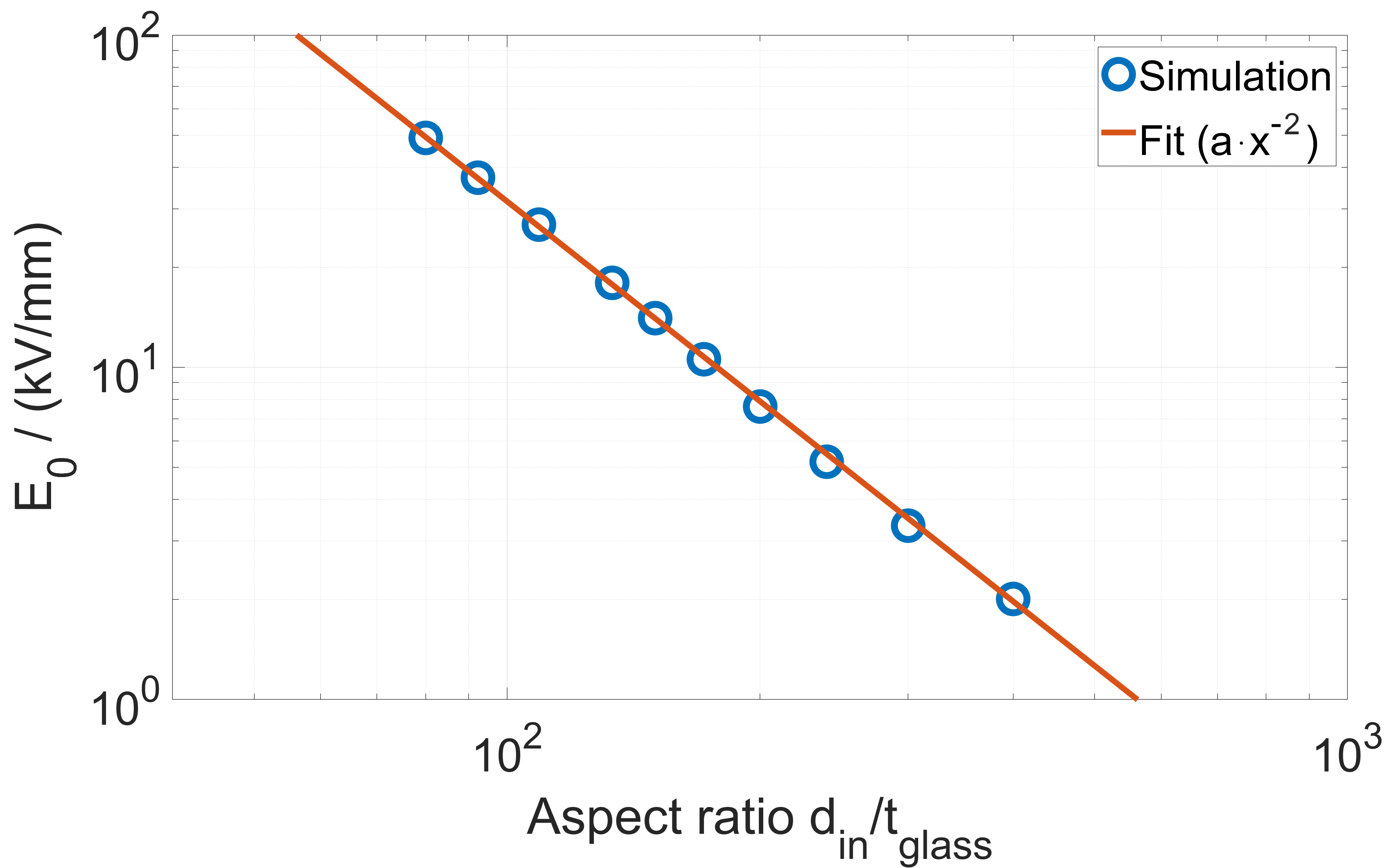}
	\includegraphics[width=0.49\linewidth]{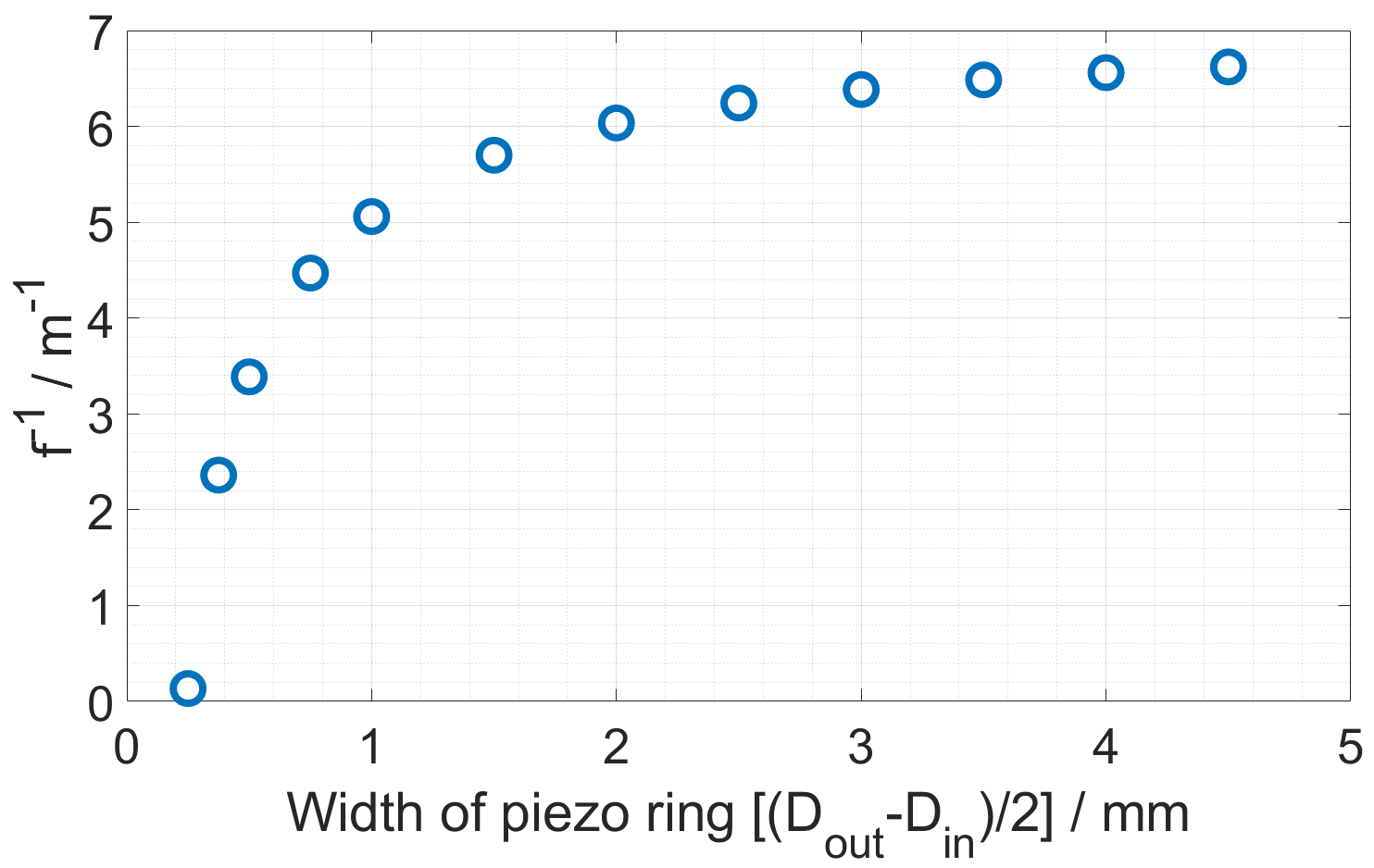}
	\caption{Left: Buckling onset $E_{0}$ as a function of the aspect ratio, simulated at a fixed piezo width with $D_{\text{out}}=\SI{18}{\milli\meter}$ and aperture $D_{\text{in}}=\SI{12}{\milli\meter}$. Right: Maximum focal power in the buckling mode as a function of the width of the piezo ring, $(D_{\text{in}}-D_{\text{in}})/2$, for a \SI{50}{\micro\meter} glass membrane and \SI{12}{\milli\meter} aperture.}
	\label{fig:dimensions}
\end{figure*}

Finally, in \fref{fig:thermalstrain} we find that a higher difference between the thermal expansion coefficients of the thin glass ($\alpha_{glass}=\SI{7.2e-6}{\per\kelvin}$) and the piezo (varied between $\SI{5.4e-6}{}$ and $\SI{7.2e-6}{\per\kelvin}$) or a higher curing temperature leads to a reduction of the focal power in the bending mode by over 30\%. Furthermore, the buckling deflection is reduced as well, in particular due to an increasing buckling threshold. As the same effect occurs also in variations of the operating temperature, it may hence be very important to find a glass/piezo combination with similar expansion coefficients.
To identify the critical dimensions combinations we simulated the buckling onset depending on the aspect ratio (\fref{fig:dimensions} left) and the focal power in the buckling mode for different piezo ring widths (\fref{fig:dimensions} right).  In both cases we ignored hysteresis and assumed a constant $d_{31}=\SI{-487e-12}{\meter\over\volt}$. We find, as expected, that the onset of buckling decreases with increasing aspect ratio as approximately $E_0 \approx \SI{3.14e5}{} \left(d/t\right)^{-2} \SI{}{\kilo\volt\over\milli\meter}$. Hence, one needs to have a minimum aspect ratio of approximately $d \gtrsim 100 \, t$. Looking at the piezo width on the right of \fref{fig:dimensions}, we see that the effect of the increasing strength of the piezo starts to saturate once the radial cross section of the piezo, $2\; t_{\text{piezo}} (D_{\text{out}}-D_{\text{in}})$, equals the radial cross section of the glass membrane $t_{\text{glass}} D_{\text{out}}$.

\section{Summary and conclusions} \label{sec:res}
We successfully demonstrated the simulation of a non-linear piezo bending and buckling actuator at the example of a varifocal adaptive lens with spherical correction. We included the non-linear charge coefficient of the piezo into our simulation, including hysteresis and creep effects, by first measuring the curvature of a well-defined cantilever actuator and then modifying the voltage in the simulation accordingly. We found that the simplified analytical approach based on geometric deformation and neglecting forces matched our measured results within a factor of 2 \myEmph{and explained possible reasons for this deviation. Nevertheless, the analytical calculations describe the qualitative scaling behavior of the lenses very well.} Comparing two lenses with identical parameters, we found a good reproducibility of the behavior, up to some variation in the pre-deflection. 

\myEmphTwo{A combination of this pre-deflection and the fabrication tolerances, such as glue layer thickness, width of the glue edge and thermal strain, are most likely the cause of the small deviations between the measurements and the simulations.}

As overall conclusions for the design, we find that in general, smaller apertures inversely proportionally increase the focal power range and thinner membranes increase the aspherical operating region. As we found that the buckling offset voltage scales quadratically with the aspect ratio of the membrane, thinner membranes also enable smaller apertures. The width of the piezo rings had relatively little effect, provided that they are wide enough to cause buckling, and there was in general little effect on the bending mode.

Furthermore, we analyzed the effects of fabrication uncertainties to develop a realistic simulation and provide the basis for optimizations of the lens. We found that changes in glue edges have a relatively small influence of the order of the reproducibility of the fabrication, that appears only in the large buckling displacement. The glue layer caused a similarly small decrease in  the focal power and aspherical tuning range in the bending mode, so one should try to minimize it, if possible. The thermal stress due to elevated curing temperatures, however, significantly affects the behavior: It greatly reduces the focal power in the bending mode and causes a small increase in the buckling onset, but also a small improvement in the aspherical tuning range at small fields. As temperature variations during operation have the same effect, it will be important to find material combinations with similar thermal expansion coefficients.

Our models now provide a reliable toolbox to design and optimize our adaptive lenses towards desired properties. In the future we aim to improve the fabrication process to reliably fabricate lenses with a membrane thickness of only \SI{30}{\micro\meter}, reduced thermal stress and minimal glue layer. To further modify the operating region, we will need to study the effects of the lens chamber, that may be used to adjust the counter pressure and the mechanics of the lens.

\section*{Acknowledgments}
This work was supported by German Research Foundation (DFG) grant WA 1657/6-1 and by the BrainLinks-BrainTools Cluster of Excellence (DFG grant EXC 1086).


\section*{References}
\bibliographystyle{unsrt}
\bibliography{Bib}
\end{document}